\documentclass[final,3p,times]{elsarticle}

\usepackage{amssymb}
\usepackage{amsmath}

\journal{International Journal of Human-Computer Studies}

\input{macros}

\begin{document}

\begin{frontmatter}



\title{Design Patterns of Human-AI Interfaces in Healthcare\tnoteref{tnote1}} 

\tnotetext[tnote1]{%
This article has been accepted for publication in the \textit{International Journal of Human--Computer Studies}. The final published version can be available at \texttt{https://doi.org/10.1016/j.ijhcs.2026.103737}.
}


\author[HKUST]{Rui Sheng}
\ead{rshengac@connect.ust.hk}

\author[SEU]{Chuhan Shi}
\ead{chuhanshi@seu.edu.cn}

\author[HKUST]{Sobhan Lotfi}
\ead{sobhan.lotfi121@gmail.com}

\author[ASU]{Shiyi Liu}
\ead{jo.liushiyi@outlook.com}

\author[CMU]{Adam Perer}
\ead{adamperer@cmu.edu}

\author[HKUST]{Huamin Qu}
\ead{huamin@cse.ust.hk}

\author[ETH]{Furui Cheng \corref{cor1}}
\ead{furui.cheng@inf.ethz.ch}

\cortext[cor1]{Corresponding author}
\affiliation[HKUST]{organization={Hong Kong University of Science and Technology},
            city={Hong Kong},
            state={Hong Kong},
            country={China}}
\affiliation[ETH]{organization={ETH Zürich},
            city={Zürich},
            state={Zürich},
            country={Switzerland}}
\affiliation[SEU]{organization={Southeast University},
            city={Nanjing},
            state={Jiangsu},
            country={China}}
\affiliation[ASU]{organization={Arizona State University},
            city={Tempe},
            state={Arizona},
            country={United States}}
\affiliation[CMU]{organization={Carnegie Mellon University},
            city={Pittsburgh},
            state={Pennsylvania},
            country={United States}}

\begin{abstract}
Human–AI interfaces play a pivotal role in integrating clinicians’ expertise with artificial intelligence to enhance both healthcare practice and research.
However, designing effective interfaces in this domain remains a significant challenge.
The inherent complexity of medical data, the influence of domain-specific conventions, and the diverse needs of clinical users compound the challenge of developing practical and usable solutions.
\sidecomment{R3C13, R1C2}
\rui{
In this study, we review existing solutions and synthesize a set of design patterns---recurring approaches that support the design of human–AI interfaces in clinical settings.
We conducted a comprehensive literature review of human–AI interaction designs in clinical contexts, through which we identified 15 information entities commonly presented to users and 12 design patterns used to organize and communicate this information effectively.
For each design pattern, we summarize the underlying design problem, the proposed solution, and the rationale for when the pattern should or should not be applied, based on insights from both the literature and semi-structured interviews with 12 healthcare professionals.
We evaluated the proposed design patterns through an online workshop involving 14 experienced UI designers. During the workshop, participants were asked to create interface sketches for healthcare-related scenarios drawn from their own professional experience, using our design patterns as guidance.
Our findings show that the proposed design patterns helped participants ground their designs in user needs, generate a wider range of design alternatives, and simplify complex interface structures.
We further analyzed and summarized the participants’ usage strategies and feedback regarding the applicability and usefulness of the design patterns.
By consolidating recurring solutions and design rationales, our work provides a practical foundation for creating more efficient and clinically meaningful human–AI interfaces, ultimately advancing the integration of AI into real-world healthcare practice.
}

\end{abstract}

\begin{graphicalabstract}
\includegraphics[width=1\textwidth]{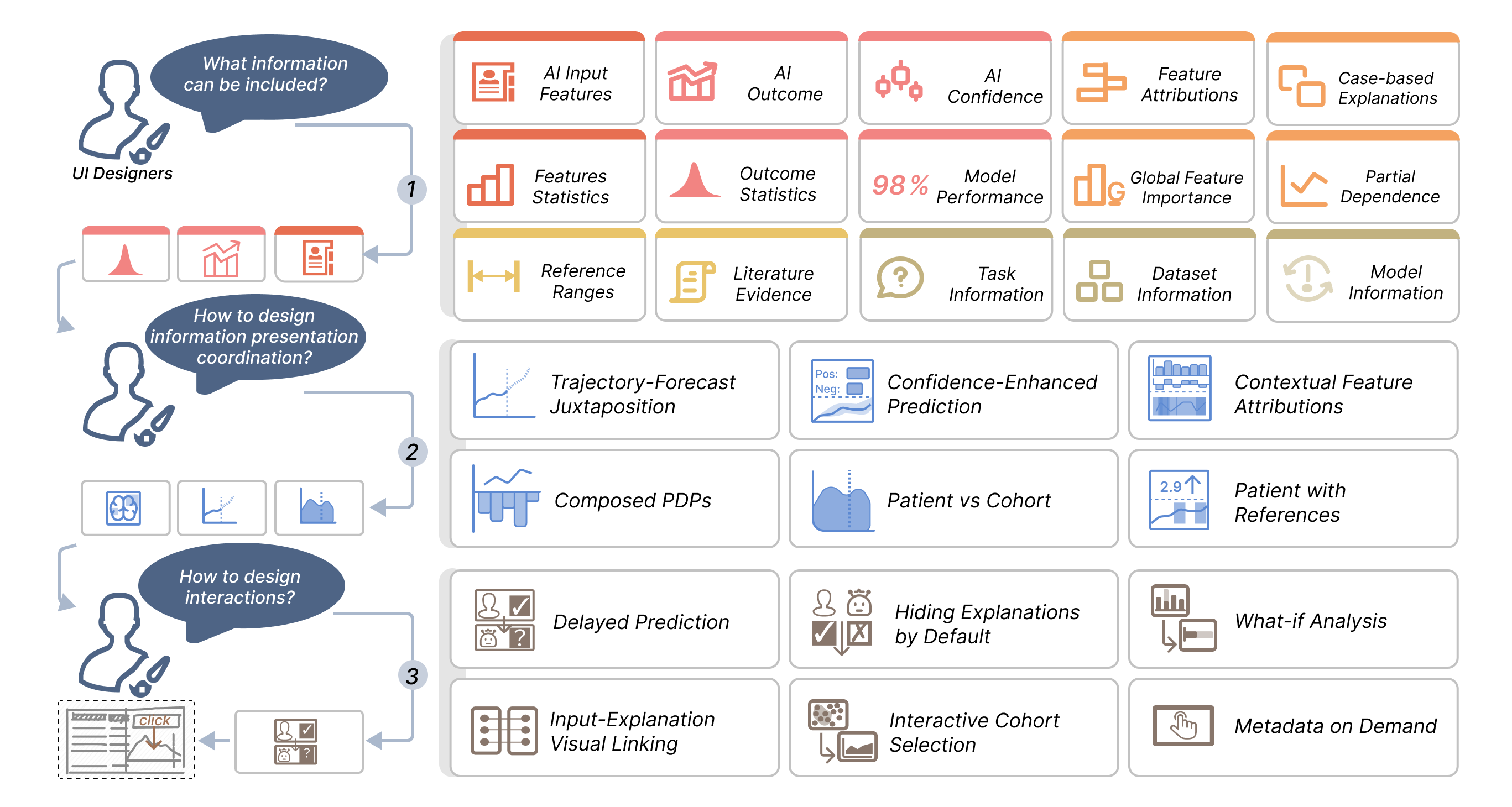}
\end{graphicalabstract}

\begin{highlights}
\item The study systematically reviews Human–AI interfaces in clinical settings and synthesizes 15 information entities and 12 reusable design patterns to address the complexity of medical data and workflows.
\item Each design pattern is grounded in both prior literature and interviews with 12 healthcare professionals, outlining the design problem, solution, and conditions for appropriate use.
\item An evaluation with 14 experienced UI designers shows that the design patterns help ground interface concepts in user needs, expand design exploration, and simplify complex interface structures.
\item By consolidating recurring solutions and rationales, the work provides a practical foundation for designing clinically meaningful Human–AI interfaces that support real-world healthcare practice.

\end{highlights}

\begin{keyword}
Design Patterns \sep Human-AI Interface \sep Healthcare
\end{keyword}

\end{frontmatter}


\section{Introduction}\
\sidecomment{R3C10}
\rui{Artificial intelligence (AI)} has emerged as a powerful tool in healthcare, with the potential to reduce healthcare-related decision-making efforts and enhance decision quality. 
This can be achieved through incorporating AI capability to forecast patients' conditions, recommend treatments, and perform various other tasks~\citep{Francisco2022Modeling, DU2024DeepThInk, GU2024Majority, JIANG2022Everyday, SEITZ2022Can, Katarzyna2023Co-designing}.
In healthcare practice, AI is integrated into users' decision-making workflows through a human-AI interface~\citep{Jadhav20223D, Lee2019Learning, Lindvall2021Rapid, Xie2020CheXplain}, which communicates AI outcomes to relevant stakeholders (\eg, clinicians, radiologists, and pathologists) and assists them in comprehending and effectively utilizing AI results to make informed decisions.

In the past few decades, the field of human-computer interaction (HCI) has focused on exploring human-AI interfaces' design and proposing general design principles and guidelines~\citep{amershi2019guidelines, mandel1997elements, shneiderman2010designing}. 
Since the 1990s, the community has been reflecting on UI design practices and proposed general principles~\citep{mandel1997elements, shneiderman2010designing}, such as Theo Mandel's three golden rules in UI design: ``\textit{place the user in control}'', ``\textit{reduce the user's memory load}'', and ``\textit{make the interface consistent}''~\citep{mandel1997elements}.
A relatively recent study by Amershi et al. summarized 18 general applicable human-AI interaction design guidelines that can be used for multiple AI application scenarios~\citep{amershi2019guidelines}.
Despite the community's efforts in developing high-level design principles and guidelines, the translation of these principles into concrete designs for healthcare-related human-AI interfaces remains a significant challenge for designers.
\sidecomment{R1C3\_1}
\rui{This challenge is further amplified in healthcare settings, where the complexity of medical data, domain-specific conventions (\eg, using patient cohorts as references for individual cases), and the diverse user needs in clinical decision-making make designing effective and clinically appropriate interfaces particularly difficult.}

In this study, we aim to provide concrete and systematic guidance for designing human-AI interfaces in healthcare by summarizing design patterns---reusable solutions for recurring design challenges.
\sidecomment{R3C1}
\rui{The concept of design patterns was proposed by Christopher Alexander et al. in architecture~\citep{alexander2018pattern}} and then generalized to more disciplines, such as software engineering~\citep{gamma1993design, heer2006software} and UI designs~\citep{tidwell2010designing}. 
Compared to design guidelines, patterns are more concrete and often include a clear description of usage context~\citep{van2001patterns}, enabling designers to quickly understand when and how to use them.

In healthcare, human-AI interfaces often communicate enriched types of information, such as the patient's demographics (as the AI input), model predictions, and explanations, which is a significant difference from general UIs. 
Designers face the challenge of selecting the proper types of information entities to present in the human-AI interface. This requires expertise in clinical data systems and (explainable) machine learning.
This design challenge motivates the first part of our study, where we aim to understand the \textit{types of information entities involved in human-AI interfaces}, 
\sidecomment{R4C4}
\rui{such as information related to AI inputs, AI-generated outputs, and AI explanations.} 
We conducted a thorough literature interview on relevant literature published in HCI venues and selected 43 research papers that have contributions to human-AI interface designs.
These papers are treated as the best practices in human-AI interface designs.
\sidecomment{R1C3\_2}
\rui{We summarized the information entities included in these papers and summarized common justifications, considering the healthcare context, to help designers understand when to use each information entity and how they relate to complex medical data (Section \ref{sec:info-entity}).}
By treating the information entities as the basic elements, we then investigated \textit{how they are visually presented in the interface}.
\sidecomment{R4C1}
\rui{Specifically, we focus on information presentation coordination approaches for different information entities and identified six patterns (Section \ref{sec:info-pre}), which describe how different information entities can be visually organized and combined within an interface.
At last, we summarized six design patterns about \textit{interaction design} (Section \ref{sec:info-org}), which specify ways in which different information entities can be dynamically interacted with or manipulated.}
\sidecomment{R1C3\_2, R1C5\_3}
\rui{These patterns take into account clinical conventions in decision-making, such as the prioritization of information and the relevance between different pieces of information. This approach helps designers create systems that support effective and reliable performance in high-risk healthcare scenarios.
}

\sidecomment{R4C2}
\rui{To ensure that these design patterns are grounded in real-world clinical practice, we conducted semi-structured interviews with 12 healthcare professionals. These interviews provided feedback on the applicability of the patterns and captured domain-specific considerations, which directly informed the usage of each design pattern.}
Then, we conducted an online workshop with 14 UI designers who have healthcare human-AI interface design experiences to understand how designers use the proposed design patterns.
In the workshop, participants were asked to select a design context from their design experiences, and then sketch a human-AI interface using Miro.
Their feedback highlights the effectiveness of the proposed design patterns in offering systematic design guidance. 
\rui{Building on these insights, our design patterns have the potential to support the design and development of human-AI interfaces by aligning with healthcare conventions, offering concrete design approaches, and highlighting critical considerations, ultimately facilitating interfaces that improve clinical decision-making performance and efficiency.}

In summary, our contributions include
\begin{itemize}[topsep=1mm]
    \item \textbf{Information entities} involved in human-AI interfaces in healthcare, which are summarized from the collected papers. These information entities can guide designers in the first design step (\ie, information selection), and they also serve as fundamental components for our design patterns related to information presentation and interaction. 
    \sidecomment{R4C3}
    \rui{Specifically, our findings not only encompass AI inputs, outputs, and explanations, but also highlight the critical role of external knowledge (\eg, literature evidence) and metadata (\eg, information about the AI model's training dataset) in supporting clinicians' decision-making.}
    
    \item \textbf{Design patterns} for information presentation coordination and interaction in designing healthcare-related human-AI interfaces. 
    Additionally, we interviewed 12 healthcare professionals to provide deep insights into these patterns, enhancing their actionable relevance.
    \sidecomment{R4C3}
    \rui{Our findings highlight common information presentation coordination patterns that can guide the design of interfaces aligning with clinicians’ established conventions—for instance, presenting a patient’s data in comparison with that of their cohort. In addition, the identified interaction strategies can help designers consider factors such as decision-making efficiency and objectivity, thereby supporting more effective and clinically appropriate human-AI collaboration.}

    \item \rui{\textbf{Insights} about user reflections and common usage strategies of our design patterns through the workshops with 14 UI designers. These insights can provide valuable guidance for healthcare human-AI interface designs in real-world applications.}
\end{itemize}
\section{Related work}
\subsection{Human-AI Interfaces in Healthcare}
The integration of AI is enhancing efficiency and personalization in the healthcare field, including both practice and research.
Numerous systems have been developed for different target users, mainly focusing on health professionals, such as clinicians~\citep{Ayobi2023Comp, Burgess2023Healthcare, Eulzer2023Fully, Jadhav2022Covid, Murray2021MedKnowts, Wang2024Surgment, yang2019unremarkable, Yang2023Harnessing}, pathologists~\citep{Cai2019Human-Centered, Gu2023Augmenting}, and radiologists~\citep{Bach2023If, Xie2020CheXplain}.
They utilize AI for tasks like diagnosis, treatment planning, and patient monitoring. 
For example, Sivaraman et al.~\citep{Sivaraman2023Ignore} developed a decision-making system that can provide treatment suggestions for sepsis. 
The system allows clinicians to make decisions with the help of AI on real patient cases. 
Zhang et al.~\citep{Zhang2024Rethinking} also leveraged AI to help clinicians predict the sepsis risk of patients.
In addition to health professionals, some systems also focus on patients~\citep{Barth2024Glimpse, Mitchell2021Reflection}. 
For instance, Mitchell et al.~\citep{Mitchell2021Reflection} introduced GlucoGoalie, leveraging AI to recommend personalized nutrition suggestions for individuals with type 1 diabetes. 
In addition, several systems target researchers to accelerate healthcare research~\citep{Floricel2022THALIS, Floricel2024Roses, jiang2023healthprism, Wang2023Extending, wang2024kmtlabeler}. 
For instance, HealthPrism~\citep{jiang2023healthprism} can help researchers examine the impact of different context and motion features on children's health.
Finally, some works also emphasize the importance of assisting ML developers in improving models~\citep{Sielaff2023Visual, wang2024kmtlabeler, Xu2023Predictive}.
Those diverse systems provide rich support for studying how to design human-AI interfaces for the healthcare domain, particularly in understanding what types of information can be effectively integrated to meet the needs of various users. 
However, there is a lack of a systematic summary of those historical design solutions.
In this work, we focus on analyzing the information entities involved in those healthcare-related human-AI interfaces.

\subsection{Information Entities in Human-AI Interfaces}
Recent studies have begun to examine the key information entities involved in human-AI interfaces~\citep{eigner2024determinants, gomez2023designing, Lai2023Towards, Momose2024Human-AI, Subramonyam2022Solving}. 
These research efforts can play a vital role in guiding the design and enhancement of such systems for designers and researchers.
However, those works usually provide general summaries of various information entities~\citep{eigner2024determinants, gomez2023designing, Momose2024Human-AI, Subramonyam2022Solving}, or approach the problem from a technical perspective~\citep{Lai2023Towards}, without thoroughly examining the considerations of the design process.
For example, Subramonyam et al.~\citep{Subramonyam2022Solving} summarized key components in AI-driven user interfaces such as input, output, and explainability, which is overly general.
Lai et al. proposed four types of information in human-AI systems, such as predictions and information about predictions, based on \textit{``technical availability''}~\citep{Lai2023Towards}.

More importantly, few studies adequately address the unique information requirements in the healthcare context. 
Given the high-stakes characteristic of this field, the healthcare domain should prioritize tailored information. 
Although there is a study~\citep{Schoonderwoerd2021Human-centered} that elaborates on the information entities of AI explanations for clinicians, this narrow scope is still insufficient to meet the diverse needs of end users such as clinicians or patients. 
For example, clinicians may require external literature evidence to judge AI predictions~\citep{Yang2023Harnessing}, following the principles of evidence-based medicine. 
Therefore, it is urgent to propose systematic guidance to help designers understand information entities involved in healthcare-related human-AI interfaces with how to organize them during the design process.

\subsection{Design Patterns}
Design patterns serve as reusable solutions to common design challenges.
In contrast to design spaces or taxonomies, they are not exclusive constructs.
Instead, they can be combined and used in conjunction with one another~\citep{Bach2023Dashboard}.
Each design pattern typically comprises four essential components: a name that identifies the pattern; a problem statement that describes the challenges indicating when to use the pattern; a solution outlining how the pattern addresses the problem; and consequences discussing potential advantages and considerations of its implementation~\citep{gamma1995design}.
There are multiple papers focused on developing design patterns~\citep{ Brehmer2017Timelines, Bach2018Comics, Bach2023Dashboard, Schoonderwoerd2021Human-centered, Tjeerd2022learning}.
For example, data-comic design patterns were distilled by analyzing multiple data videos, infographics, and other existing data comics~\citep{Bach2018Comics}.
In addition, three dimensions, each containing specific patterns, have been proposed for the creation of expressive data stories~\citep{Brehmer2017Timelines}.
The closest work to ours is the dashboard design patterns~\citep{Bach2023Dashboard}. 
It outlined the information entities (\eg, single values, derived values) to consider when designing a dashboard and how to present them. 
Additionally, it discussed how to organize these information entities on interfaces, such as page layout and screen space.
However, it does not take into account the unique aspects of integrating AI into interfaces.

\section{Methodology}
\sidecomment{R4C5, R3C7\_3, R1C5\_2}
\rui{We followed the guidelines of Kitchenham's systematic literature review protocol~\citep{kitchenham2004procedures} to ensure transparency and reproducibility. Our process consisted of the main steps that correspond to key stages in Kitchenham’s framework: (1) paper identification and selection, (2) data extraction and synthesis. 
During the paper identification stage, one author collected papers from targeted venues using predefined keywords. Subsequently, three authors conducted the screening of these papers based on inclusion and exclusion criteria.
For data extraction and synthesis, three authors independently performed the coding. Any discrepancies were resolved through group discussion.
Finally, to ensure that these design patterns are rooted in real-world clinical practice, we conducted semi-structured interviews with 12 healthcare professionals regarding their usage examples, advantages, and considerations.
}

\begin{figure}
    \centering
    \includegraphics[width=1\linewidth]{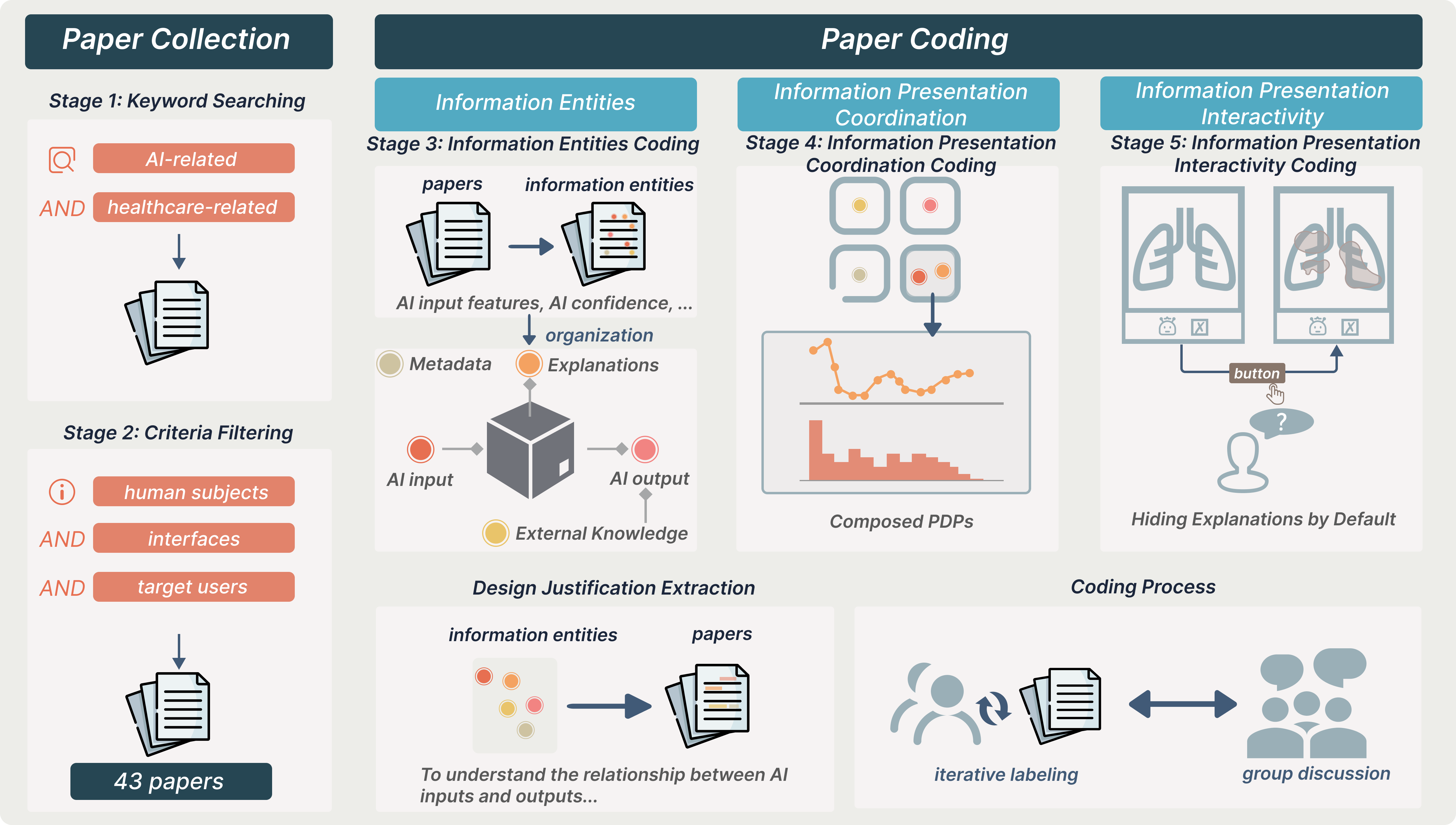}
    \caption{The paper collection and coding process.}
    \label{methodology} 
\end{figure}

\subsection{Paper Identification and Selection}
We commenced our work by reviewing relevant research on human-AI interfaces in healthcare, as we aimed to derive design patterns with their corresponding justifications mentioned in the literature.
We searched for papers published in various venues, including five kinds of journals (\ie, IJHCS, TVCG, ToCHI, IJHCI, THMS) and eleven types of conferences (\ie, CHI, EuroVis, IUI, UIST, FAcct, PacificVis, PACMHCI, DIS, HEALTH, KDD, and CSCW), over the last ten years, from 2015 to 2024. 
\sidecomment{R3C7\_1, R3C7\_2}
\rui{To ensure the quality of the included papers, most journals and conferences were selected based on their inclusion in the CCF (China Computer Federation) ranking. A few venues such as FAccT and PacificVis, although not formally ranked by CCF, were included due to their high visibility in interdisciplinary research and their appearance in previous survey papers~\citep{wu2022ai4vis, Li2024where, tahaei2023systematic}. 
Searches were conducted using the most commonly used engines for each venue: IEEE Xplore for TVCG and THMS, ACM Digital Library for CHI, IUI, UIST, KDD, HEALTH, CSCW, FAccT, ToCHI, PACMHCI, and DIS, and the corresponding official conference or journal websites for other venues.
}
First, we collected all the relevant papers by keyword searching. 
There are two types of keywords: (1) healthcare-related keywords (\eg, clinical, patient, disease, diagnosis, etc) refined from a previous paper~\citep{Tran2019Global}; and (2) AI-related keywords (\eg, machine learning, deep learning, etc) summarized by our team.
\sidecomment{R4C5}
\rui{We obtained 2,168 papers at this time.}
Further, we checked whether these papers were in our scope based on two criteria. 
\begin{itemize}
    \item The paper proposes and justifies a human-AI interface or system to solve a healthcare-related task. We excluded papers that only implement a conventional human-AI interface design without providing design justifications. 
    \item The target users of the proposed system are health professionals (\eg, clinicians or radiologists). We excluded those systems designed for patients~\citep{Liang2020OralCam, Barth2024Glimpse}, AI developers~\citep{Sielaff2024Bayesian}, and biomedical researchers~\citep{jiang2023healthprism, Wang2023Extending} as these systems share very different design objectives with health-professionals-oriented systems.
\end{itemize}
To confirm a paper's relevance, we first reviewed the title and abstract, then checked the other sections if needed.
Finally, we obtained 43 papers.

\subsection{Data Extraction and Synthesis}
We followed the three-layer structure (information selection, information presentation coordination, and interaction design) to identify the design patterns in the collected papers (\autoref{methodology}).

We first labeled the information entities in the interfaces and their corresponding design justifications in the papers. 
\rui{The labeling schemes were then refined iteratively through regular meetings held over the course of one month.}
We consider the meaning of the information entities in the AI pipeline and their interpretations to the target users. 
For example, a patient's lab test results can be interpreted as AI input features in an AI pipeline. 
To create a more general and systematic taxonomy, we consider the information entity's AI meaning (\eg, input features) as the primary category and its medical interpretation (\eg, lab tests) as a complementary property. 
When coding the design justifications, we mainly refer to the formative study, design requirements, and system design sections. 

In the next step, we coded the presentations of the information entities identified from the previous step.
In this work, we focus on information presentation coordination, which refers to how different information entities can be presented together cohesively.
\sidecomment{R4C5}
\rui{We only coded designs that were accompanied by explicit design justifications, ensuring that the extracted design patterns were grounded in the rationale provided by the authors.}

Finally, we labeled the interaction design, which encompasses the order of information presentations and the interactive coordination approaches between different information entities.
We use a similar methodology to the second step, where we focus on the designs with clear justifications mentioned in the paper. 
The final coding result can be checked\footnote{https://canary-wormhole-ef5.notion.site/Paper-Coding-15fe0511aa9081ac821df741649740d7?pvs=4}.

\subsection{Scenarios of the Collected Papers}

\begin{figure}
    \centering
    \includegraphics[width=0.8\linewidth]{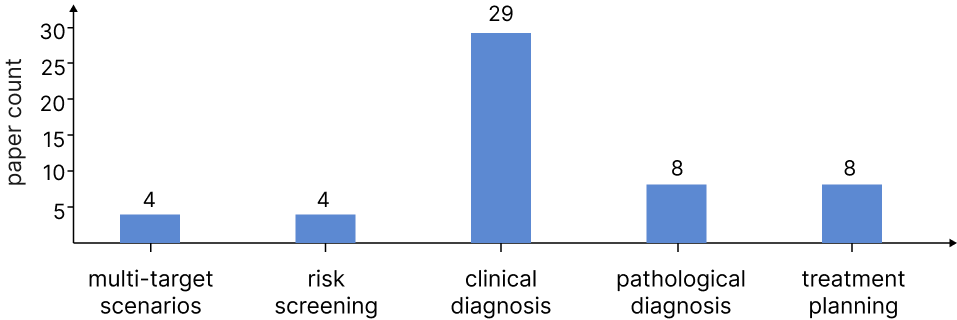}
    \caption{\rui{The distribution of scenarios in the collected papers.}}
    \label{count} 
\end{figure}

We analyzed 43 papers and identified four key scenarios: risk screening, clinical diagnosis, pathological diagnosis, and treatment planning. Risk screening (4/43) involves testing asymptomatic individuals for diseases, typically conducted by clinicians. Clinical diagnosis (29/43) refers to clinicians diagnosing patients based on experience, lab tests, and observational data. Pathological diagnosis (8/43) involves pathologists diagnosing through tissue scanning, while treatment planning (8/43) entails selecting appropriate interventions to manage diseases, also usually performed by clinicians. Most papers (39/43) focused on a single scenario, though some (4/43) addressed multiple ones, such as CarePre~\citep{Jin2020CarePre}, which utilized patient EHR data for risk prediction and treatment exploration.

\subsection{Semi-structured Interview}
\begin{table}[h]
\centering
\caption{Demographics of the 12 health professionals interviewed, including their specialty, years of practice, and gender.}
\label{table1}
\begin{tabular}{l|l|l|l}
\hline
\textbf{ID} & \textbf{Specialty}       & \textbf{Years of practice} & \textbf{Gender} \\ \hline
E1          & Dermatology               & 3 years             & Male            \\ \hline
E2          & Orthopedics               & 7 years             & Male            \\ \hline
E3          & Nephrology                & 1 year              & Female          \\ \hline
E4          & Dermatology               & 3 years             & Female          \\ \hline
E5          & Obstetrics and Gynecology & 15 years            & Female          \\ \hline
E6          & Gastroenterology          & 3 years             & Female          \\ \hline
E7          & Oncology                  & 10 years            & Male            \\ \hline
E8          & Oncology                  & 2 years             & Male            \\ \hline
E9          & Nephrology                & 5 years             & Female          \\ \hline
E10         & Orthopedics               & 2 years             & Male            \\ \hline
E11         & Nephrology                & 30 years            & Male            \\ \hline
E12         & Nephrology                & 20 years            & Male            \\ \hline
\end{tabular}
\end{table}
After coding the collected papers, we obtained 15 information entities and 12 design patterns for information presentation and interaction (\ie, 6 information presentation coordination patterns and 6 interaction design patterns).
\sidecomment{R4C6}
\rui{We noted that the justifications for certain design patterns were primarily based on the designers' own perspectives rather than being user-centric. To further evaluate these design patterns from the end users' point of view, we conducted interviews with 12 healthcare professionals (E1-E12) (\autoref{table1}).}
For each interview, we began by introducing our background. Following this, we presented 15 information entities and 12 design patterns along with relevant examples.
\sidecomment{R4C6}
\rui{Then, for each design pattern, participants were asked open-ended questions about scenarios in which they might apply them based on their experience, along with their potential advantages and any considerations that should be noted.}
The entire process lasted approximately one hour for each interview. After the interview, they were given a small monetary reward as a gesture of appreciation.
\sidecomment{R4C6}
\rui{Next, we employed thematic analysis to systematically examine the interview transcripts in order to identify common advantages and considerations associated with the design patterns. The thematic analysis was conducted by three authors, who first coded the transcripts independently. Afterward, they convened to discuss their findings and reach a consensus on the interpretations of the advantages and considerations mentioned by the healthcare professionals in the interviews.
The insights gathered from these interviews—concerning usage scenarios, advantages, and considerations not mentioned in the literature—were utilized to enhance the 12 design patterns in \autoref{sec:info-pre} and \autoref{sec:info-org}.
}

\section{Information Entities}
\label{sec:info-entity}

\begin{figure}
    \centering
    \includegraphics[width=\linewidth]{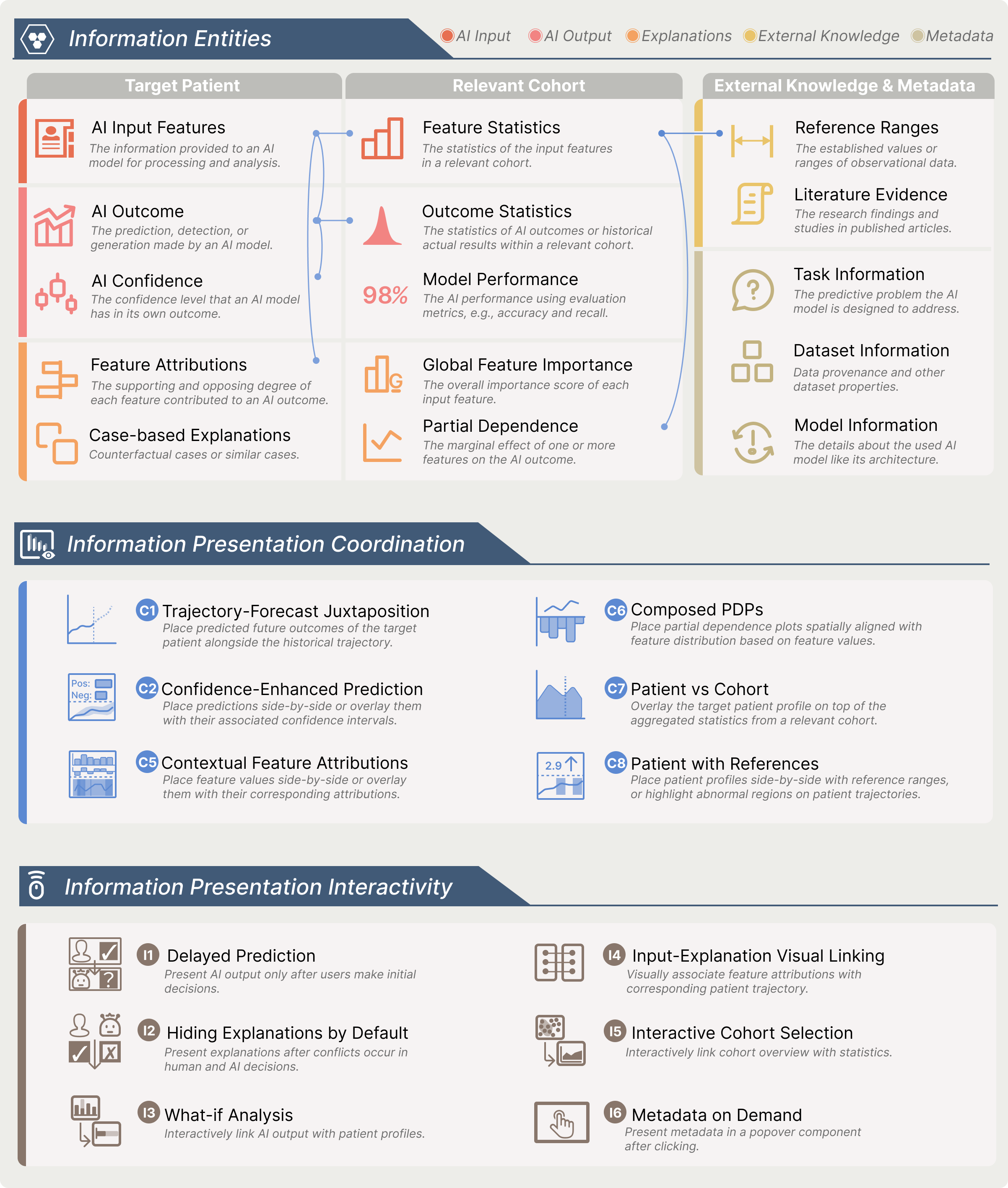}
    \caption{We identified design patterns of human-AI interfaces in healthcare in information entity selection, information entity presentation, and interaction design. 
    }
    \label{design_patterns} 
\end{figure}

Human-AI interfaces communicate enriched information entities, including patient demographics, model predictions, and explanations, making it challenging for designers to select appropriate entities. To provide more systematic guidelines, we consider both the meaning of these entities in the AI pipeline and their interpretations by target users to develop a taxonomy. In this section, we detail the information entities we have identified and their implications in the health domain, assisting designers in information selection for human-AI interface design.

\subsection{AI Input} 
AI input is the information provided to an AI model for processing and analysis. We have identified two types of information entities related to AI input.
The first is the AI input features of the target patient. The second is the feature statistics of a relevant cohort.

\pattern{icon/i1}{AI input features} refer to the various attributes of the target patient that are fed into the model.
In the healthcare domain, AI input features include patient demographics, medical events, lab tests, clinical signs, historical diagnosis or treatment, clinical notes or documents, medical images, and statistical aggregation of the patient’s medical records.
The choice of AI input features mainly depends on the task and the available dataset. 
For instance, CheXplain~\citep{Xie2020CheXplain} utilized medical images for chest X-ray analysis.
Zhang et al.~\citep{Zhang2024Rethinking} used temporal sequences of lab tests and vital signs to predict changes in sepsis risk.

\pattern{icon/i6}{Feature statistics} refer to the distribution or other statistical information (\eg, mean and standard deviation) of the input features in a relevant cohort. 
In the healthcare domain, feature statistics provide essential reference information derived from a relevant cohort, aiding stakeholders in better interpreting and comparing patient data.
For example, since aggregation features (\eg, standard deviation) are \textit{``unfamiliar to clinicians''}~\citep{Cheng2022Vbridge}, a 95\% confidence interval of feature values from a low-risk cohort was calculated as a reference.
Furthermore, when users examine AI explanations, feature statistics can be integrated to provide additional context~\citep{Krause2016Interacting}.

\subsection{AI Output}
AI output refers to the results generated by an AI model. We have identified AI outcomes and AI confidence specific to a target patient. Additionally, we can also calculate outcome statistics and assess model performance for a relevant patient cohort.

\pattern{icon/i2}{AI outcome} refers to the prediction, detection, or generation made by an AI model.
In the healthcare domain, prediction usually refers to diagnosis results or treatment types, as well as continuous values including lab tests, clinical signs, and prognostic indicators (\eg, disease risk, survival rate, and remission probabilities).
For example, Cheng et al.~\citep{Cheng2022Vbridge} aimed to inform clinicians about a patient's high risk for a specific disease, with their system providing classification predictions as output.

\pattern{icon/i3}{AI confidence} refers to the confidence level that an AI-driven system has in its own outcome. 
When users need to understand the model's capability for a given task or require additional context to interpret the AI output, AI confidence can be provided.
For instance, Lee et al. used AI confidence since users wanted to understand \textit{``how well the system can do''}~\citep{Lee2022Towards}.

\pattern{icon/i7}{Outcome statistics} refer to the statistics (\eg, distribution) of AI outcomes or actual outcomes within a relevant cohort. Similar to feature statistics, outcome statistics can serve as a historical reference or current context for users in the healthcare domain.
For example, with a predicted dropout probability from an AI model, clinicians may not know whether it is high or low \textit{``without knowing the distribution of predictions across all patients''}~\citep{Jacobs2021Design}.
In addition, Sivaraman et al. provided statistics of real-world clinician actions for sepsis treatments since users need to \textit{``compare model predictions and clinician actions''}~\citep{Sivaraman2023Ignore} to inform their decisions.

\pattern{icon/i8}{Model performance} refers to the accuracy, recall, precision, and other evaluation metrics of an AI model.
Many systems have provided model performance to help users better understand the effectiveness of an AI model~\citep{Lee2020Co-Design, Lee2021Rehabilitation}.

\subsection{Explanations}
Explanations refer to the information used to display the AI mechanism. There are two kinds of explanations. The first is instance-level explanations (\ie, feature attributions and case-based explanations), which can provide insights into why the AI produces a specific output. The second is cohort-level explanations (\ie, global feature importance and partial dependence), which can provide an overall understanding of AI's mechanism based on analyzing multiple instances. 

\pattern{icon/i4}{Feature attributions} are the supporting and opposing degree of each feature contributed to an AI outcome.
In the healthcare domain, feature attributions can indicate the contributions of different items in lab tests, clinical signs, or other observational data to the result. 
When users need to understand why an AI makes a prediction or seek explanations as decision confirmation, feature attributions can be used ~\citep{Xie2020CheXplain, Krause2016Interacting, Gu2023Augmenting, Jadhav2022Covid}.
For instance, when users need to understand \textit{``the relationship between inputs (patient records) and AI outputs (prediction scores)''}~\citep{Kwon2019RetainVis}, feature attributions can be helpful.
Furthermore, it also can be leveraged when users hope to know the importance of different features to a prediction.
For example, Krause et al. allowed users to understand \textit{``the most important features for a given patient''}~\citep{Krause2016Interacting}.

\pattern{icon/i5}{Case-based explanations} refer to using cases from the past to aid in understanding. There are mainly two types of case-based explanations. The first type includes instances where the AI outcome or confidence differs from the target (\ie, contrastive and counterfactual cases). The second type consists of instances with feature values that are similar to those of the target (\ie, similar cases).
In the healthcare domain, case-based explanations usually represent (1) cases with different diagnosis outcomes, or (2) cases with similar lab tests, clinical signs, or other observational data.
Similar to feature attributions, when users seek to understand the reason behind an AI's prediction, case-based explanations can be used~\citep{Xie2020CheXplain, Krause2016Interacting}.
Specifically, when users desire to leverage real cases as their decision-making \textit{``evidence''}~\citep{Xie2020CheXplain}, case-based explanations can be helpful.
For instance, CheXplain~\citep{Xie2020CheXplain}  displayed counterfactuals since physicians stated that comparing normal and abnormal chest X-ray images to make a decision was a common approach.

\pattern{icon/i9}{Global feature importance} represents the overall importance score of each input feature for a given AI model.
In the healthcare domain, it can indicate the importance of different items in lab tests, clinical signs, or other observational data.
Global feature importance can help users understand the model.
Specifically, when users need to identify the key features in a task, this information entity can be leveraged.
For instance, Dmitriev et al. showed global feature importance to indicate \textit{``which input features are the most important''}~\citep{Dmitriev2021Lesions} for pancreatic lesions diagnosis.

\pattern{icon/i10}{Partial dependence} represents the marginal effect of one or more features on the AI outcome while holding the values of other features constant.
In the healthcare context, it illustrates how varying values of a specific item (\eg, lab tests, vital signs) can influence the diagnosis or treatment result.
Partial dependence can be used when users hope to understand the relationship between inputs and outputs~\citep{Krause2016Interacting, Dmitriev2021Lesions}, similar to feature attributions.
For example, it has been leveraged to help users understand the impact a feature has on actual patients~\citep{Krause2016Interacting}, or \textit{``how changing the feature values''}~\citep{Jadhav20223D} influences the AI prediction.
However, compared to feature attributions, it emphasizes the overall interaction of a particular feature with the AI prediction, rather than the individual contribution of each feature to a specific prediction.

\subsection{External Knowledge}
External knowledge refers to information that originates outside the AI model or relevant dataset, including reference ranges and literature evidence. It is also a crucial component in healthcare-related decision-making with AI.

\pattern{icon/i11}{Reference ranges} refers to established values or ranges for lab test results or observational data, helping determine if a result is normal or abnormal.
For example, Zhang et al. presented the reference features to help clinicians quickly identify anomalies in patients' lab tests and vital signs~\citep{Zhang2024Rethinking}.

\pattern{icon/i12}{Literature evidence} refers to the research findings and studies in published articles, which can provide support or validation for specific practices in the healthcare domain.
It is important when users need to inspect authoritative evidence as a reference to AI output.
For instance, Yang et al. mentioned that clinicians always \textit{``shared a list of evidence from biomedical literature''}~\citep{Yang2023Harnessing} when they exchanged suggestions with peers.
This can support their practice of evidence-based medicine.

\subsection{Metadata}
Metadata refers to the descriptive information about the task, datasets, and models involved in a human-AI interface. This is usually used to provide context to help users better understand AI's capabilities and limitations. We identified three common metadata types as the following.

\pattern{icon/i13}{Task information} refers to details about the specific problem that an AI is designed to address. There have been numerous works to present task information to help users understand the functionality of systems~\citep{Francisco2021Breast, Jacobs2021Design, Wentzel2025DITTO}.

\pattern{icon/i14}{Dataset information} encompasses data provenance and other dataset properties, such as its size and sources. Several systems provided this information to help examine the data used, ensuring transparency~\citep{Jin2020CarePre, Krause2016Interacting, Ming2019RuleMatrix}.

\pattern{icon/i15}{Model information} encompasses details about the used AI model such as its architecture, hyperparameters, and training process. For example, Prospector allowed users to select different models and compare their performance~\citep{Krause2016Interacting, Ouyang2023Leveraging}.

\section{Information Presentation Coordination}
\label{sec:info-pre}
Information presentation coordination involves visually displaying multiple information entities by integrating their individual visualizations within a single view, allowing users to seamlessly interpret various information entities together.
We identified six information presentation coordination patterns.
Based on the different types of information entities, we categorized the extracted patterns into four classes: (1) target patient analysis; (2) relevant cohort exploration; (3) comparison of the target and relevant cohort; and (4) incorporation of external knowledge or metadata.

\subsection{Target Patient Analysis}
\pattern{icon/p1}{Trajectory-Forecast Juxtaposition}

\noindent
$\blacklozenge$ \textit{The problem:} 
When a forecasting model predicts the patient's future health condition based on the patient's historical data, health professionals want to compare the past trajectories with future predictions or make sense of the general trend of the patient's conditions.
The design problem is coordinating the presentations of the patient's historical data (i.e., trajectory) with the forecasting results. 

\begin{figure}[!htb]
    \centering
    \includegraphics[width=0.5\linewidth]{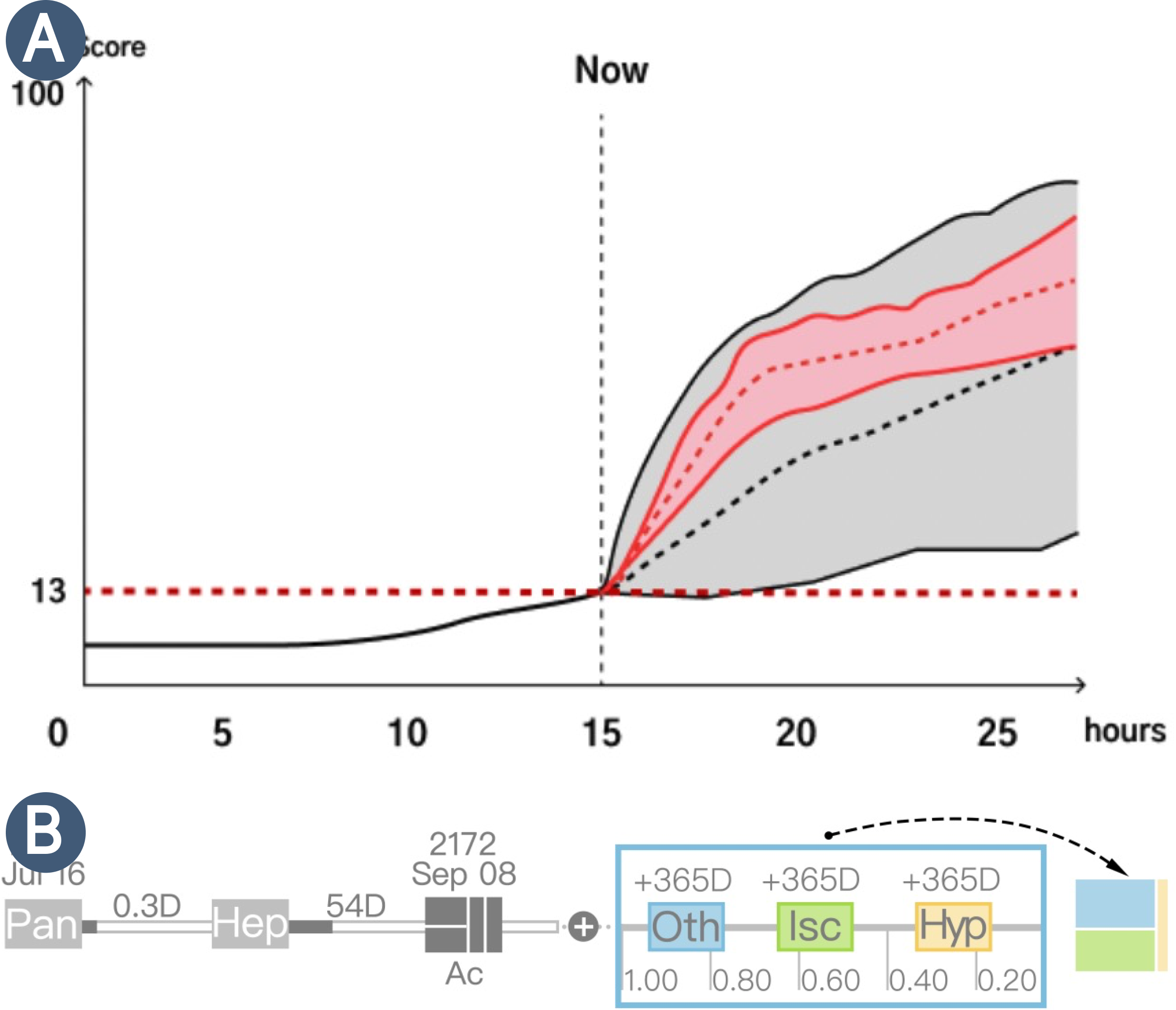}
    \caption{(A) Zhang et al.~\citep{Zhang2024Rethinking} present patient historical risk levels alongside the predicted risks. (B) Jin et al.~\citep{Jin2020CarePre} display a patient's past diseases alongside potential future diseases.}
    \label{Pattern1}
\end{figure}

\noindent
$\blacklozenge$ \textit{The solution:} When both the patient trajectory and AI forecasts are time series data, the historical trajectory and forecasting results are visualized in a single chart using juxtaposition. 
\autoref{Pattern1}-A shows an example of this pattern that composes a line chart and an area chart to visualize the patient's historical risk score (less than 15 hours) and the forecasted risk scores with their confidence intervals. 
When the patient's historical records and forecast are described using discrete medical events, they could be visualized using a single timeline (\autoref{Pattern1}-B). 

\noindent
$\blacklozenge$ \textit{Justifications:}
This pattern has been used by several existing studies when targeting a forecasting problem~\citep{Jin2020CarePre, yang2019unremarkable, Burgess2023Healthcare} and commented as ``\textit{easy to understand}'' by our participants.
Juxtaposing patient trajectories and forecasting results ensures visual consistency. In this context, these two information entities essentially represent the same data type, reflecting different time points. Making these two entities visually similar and presenting them together is crucial for establishing clear connections between them.
For instance, CarePre stated this approach could help clinicians better accomplish forecast \textit{``interpretation''}~\citep{Jin2020CarePre}.

\pattern{icon/p2}{Confidence-Enhanced Prediction}

\noindent
$\blacklozenge$ \textit{The problem:} When a forecasting model generates a prediction, health professionals want to consider the AI confidence levels in its predictions when making judgments in high-risk healthcare scenarios. The design problem is how to present predictions with AI confidence jointly.

\noindent
$\blacklozenge$ \textit{The solution:} Considering that the AI confidence level serves as additional contextual data for the AI predictions and is dependent on those predictions, designers can place predictions side-by-side or overlay them with their associated confidence intervals to illustrate the uncertainty. For example, DITTO~\citep{Wentzel2025DITTO} overlays survival prediction scores over time with an area chart to illustrate the prediction confidence.


\noindent
$\blacklozenge$ \textit{Justifications:} 
This pattern has been used in various scenarios such as risk screening~\citep{yang2019unremarkable}, clinical diagnosis~\citep{Gattupall2017CogniLearn, yang2019unremarkable, Xie2020CheXplain, goh2020ai, Jadhav2022Covid, Jadhav20223D, Zhang2024Rethinking}, treatment planning~\citep{Wentzel2025DITTO}, and pathological diagnosis~\citep{Gu2023Augmenting}.
Considering that AI confidence provides contextual information for predictions, these two information entities are usually displayed together, which can help \textit{``balance user expectations of the model''}~\citep{Jin2020CarePre}.
For time series predictions, overlaying is a common method to highlight their one-to-one relationship, helping users maintain focus. For single predictions, displaying the confidence level alongside the prediction clarifies the information.
This pattern has also been regarded as important by our participants. 
For example, one participant (E5) said, \textit{``If I ask other clinicians, they would share their opinions along with their level of confidence. This pattern effectively showcases similar information.''}

\pattern{icon/p5}{Contextual Feature Attributions}

\noindent
$\blacklozenge$ \textit{The problem:} Feature attributions demonstrate the extent to which different features contribute to the AI prediction outcomes. Considering abundant features in the health domain, health professionals sometimes want to identify salient features with the assistance of feature attributions. The design problem is displaying feature attributions to provide contextual information for feature values.

\begin{figure}[!htb]
    \centering
    \includegraphics[width=0.8\linewidth]{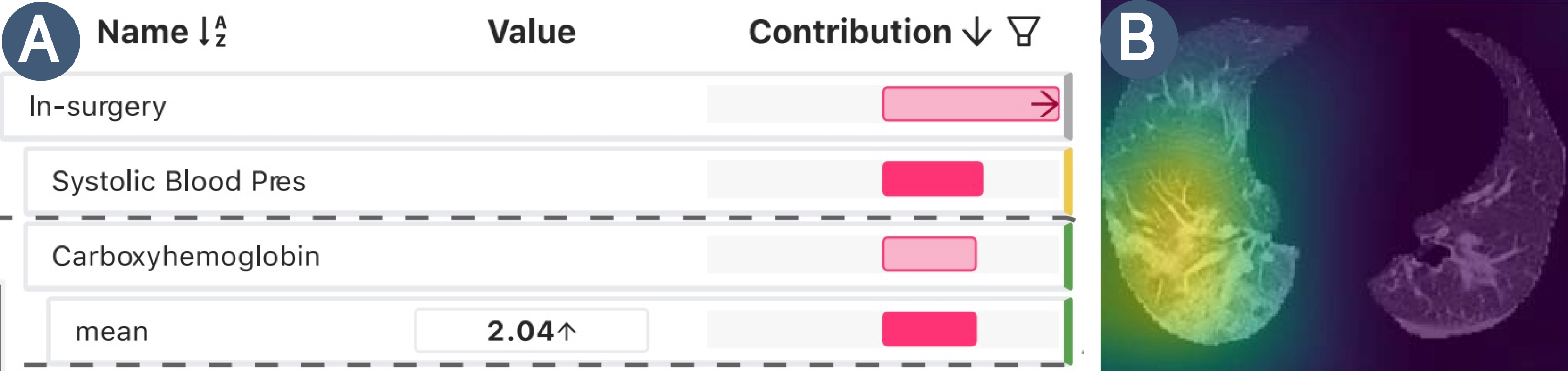}
    \caption{(A) VBridge~\citep{Cheng2022Vbridge} displays the attribution of each feature, while directly presenting the corresponding values alongside. (B) COVID-view~\citep{Jadhav2022Covid} displays Chest CT images overlaid with feature attributions.}
    \label{Pattern3}
\end{figure}

\noindent
$\blacklozenge$ \textit{The solution:} Considering that feature values and feature attributions correspond one-to-one, designers can either place the feature values side-by-side with their corresponding attributions or overlay them. \autoref{Pattern3}-A implements this pattern by using a bar chart to display the attribution of each feature, while also presenting the corresponding values alongside. 
Another example (\autoref{Pattern3}-B) overlays Chest CT images with AI-generated saliency maps, with brighter colors indicating regions that require greater attention.

\noindent
$\blacklozenge$ \textit{Justifications:} This pattern has been widely used in existing systems~\citep{Krause2016Interacting, goh2020ai, Cheng2022Vbridge, Jadhav2022Covid, Gu2023Augmenting, Gu2023Improving, Horák2023xOpat}.
This presentation method offers a convenient switch between these two information entities for users. 
Specifically, once they browse the feature attributions and identify the most important feature, they can immediately turn to view its values through spatial linking.
Similarly, users can also leverage their expertise to first identify the most noteworthy features and then grasp the corresponding feature attributions as confirmation.
However, when designers utilize this pattern---especially when overlaying these two pieces of information---it is suggested to consider the potential for visual confusion since these two types of data may become intertwined (E3 and E6). To address this, designers can offer users the option to choose whether to display the feature attributions.

\subsection{Relevant Cohort Exploration}
\pattern{icon/p6}{Composed PDPs}

\noindent
$\blacklozenge$ \textit{The problem:}
Partial dependence can illustrate how varying values of a specific feature (\ie, lab tests, vital signs) can influence diagnosis or treatment outcomes.
Health professionals need to identify whether the partial dependence is reliable or whether the current data processing methods (\eg, missing data imputation) could lead to unreasonable partial dependence. The design problem involves linking the presentation of partial dependence and feature statistics.

\begin{figure}[!htb]
    \centering
    \includegraphics[width=0.5\linewidth]{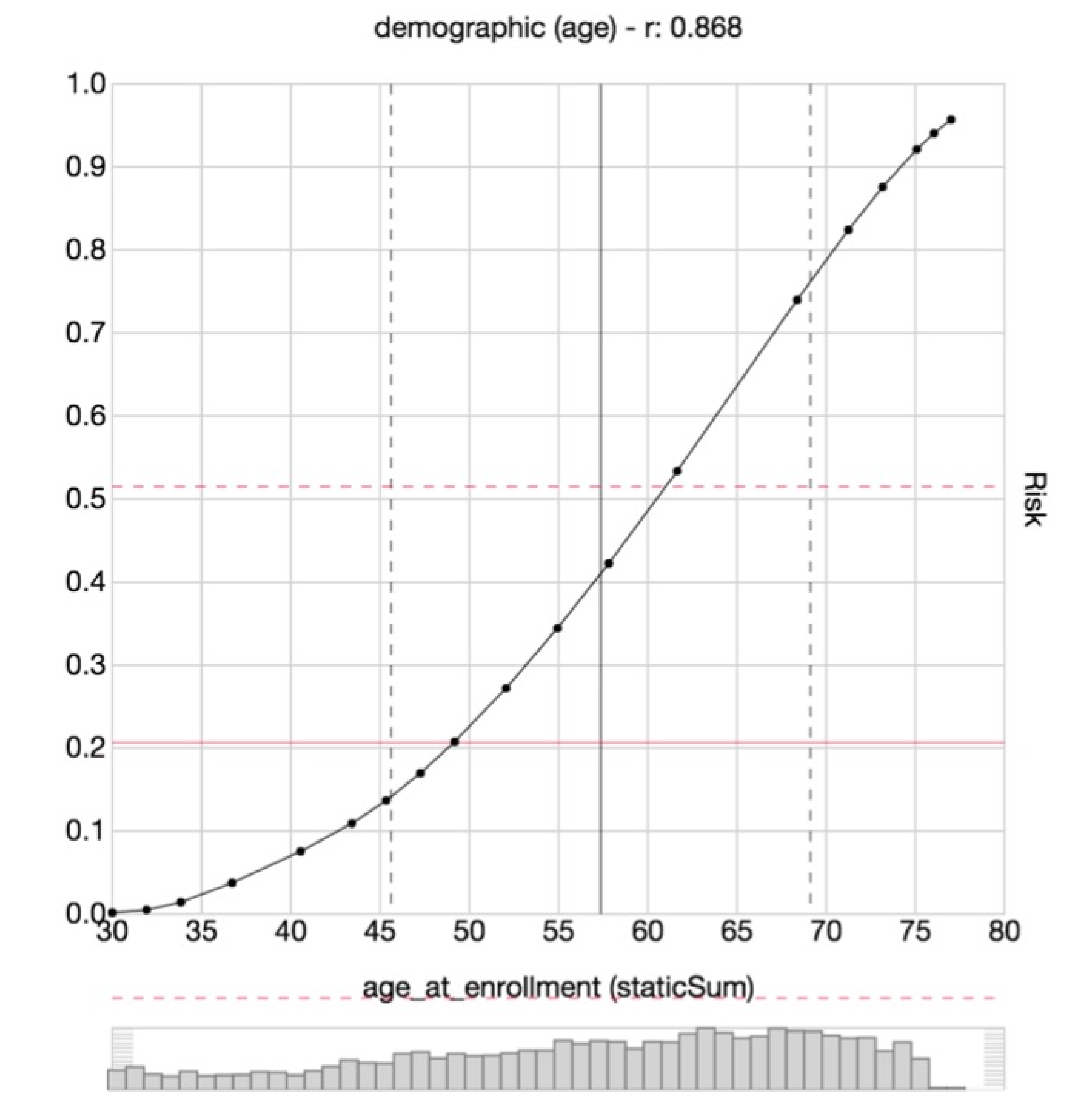}
    \caption{Krause et al.~\citep{Krause2016Interacting} place partial dependence using a line chart, while the corresponding feature distribution is displayed below using a bar chart.}
    \label{Pattern4}
\end{figure}

\noindent
$\blacklozenge$ \textit{The solution:} Designers can display partial dependence plots next to the input feature distribution. To be more specific, the visual presentations of these two information entities can be aligned based on feature values---one displaying the corresponding prediction score for that feature value (partial dependence), and the other showing the number of data samples (feature statistics) associated with it. \autoref{Pattern4} shows an example of this pattern, with partial dependence represented at the top using a line chart, while the corresponding feature distribution is displayed below as a bar chart. 

\noindent
$\blacklozenge$ \textit{Justifications:} This information provides users with more context to help them decide whether to trust the model's results. 
For instance, Propestor stated that the feature distribution below could help clinicians understand whether the model is \textit{``overfitting''}~\citep{Krause2016Interacting}.
This side-by-side arrangement enables users to quickly switch between two corresponding dimensions of a feature value: its prediction score and the corresponding number of data samples.
This pattern caught the interest of all participants. For instance, one participant (E8) highlighted, \textit{``It would be interesting to see how the AI's confidence in diagnosing a tumor changes with variations in tissue image morphology. This could enhance my diagnostic knowledge, and the accompanying training data distribution would indicate areas where I might have less trust.''}
However, when using this pattern, it is important to note that it may require users to possess sufficient AI literacy; otherwise, they may struggle to interpret the relationship between the feature distribution and the partial dependence plot (E11 and E12).

\subsection{Comparison of the Target and Relevant Cohort}
\pattern{icon/p7}{Patient vs Cohort}

\noindent
$\blacklozenge$ \textit{The problem:} 
Sometimes, the features (\eg, aggregation features like standard deviation) may undergo feature engineering, which can pose a challenge for users to understand. Similarly, the AI-predicted values may be unfamiliar to health professionals, as they may not know whether the predicted values are considered high or low. In such cases, they need to understand specific patient features or AI prediction scores, using cohort-level statistics as a reference. The design problem is displaying cohort statistics as a reference for the target patient data.

\noindent
$\blacklozenge$ \textit{The solution:} To help users better utilize cohort statistics as a reference, designers can overlay an individual patient's data on top of aggregated statistics from a relevant cohort. For example, Jacobs et al.~\citep{gomez2023designing} overlaid a specific patient's dropout prediction value onto a histogram of predictions for a relevant patient cohort.


\noindent
$\blacklozenge$ \textit{Justifications:} This pattern has been implemented in multiple existing systems~\citep{Jin2020CarePre, Cheng2022Vbridge, Jacobs2021Design, Sivaraman2023Ignore, Wentzel2025DITTO}. 
Overlaying target patient data onto the corresponding cohort statistics can achieve \textit{``spatial linking''}~\citep{Javed2012Composition}, enabling quick comparisons without requiring users to switch between different views.
In addition, this approach also saves space in the interface, especially considering that many pieces of information about patients often need to be displayed during the patient care process (E11).
Furthermore, some participants (E2, E3, E6, E7, and E12) mentioned that this design could clearly \textit{``visualize''} their clinical experience as reflected by the patient cohort, thereby providing them with greater confidence in their judgments.
However, it is important to note that the selection of the cohort is crucial, as improper choices may introduce bias (E3, E6, E10, and E12).

\subsection{Incorporation of External Knowledge or Metadata}
\pattern{icon/p8}{Patient with References}

\noindent
$\blacklozenge$ \textit{The problem:} When health professionals review patient profiles, it is challenging for them to remember the normal ranges for all the features. Therefore, they need an efficient way to identify which features fall outside the normal range. The design problem is coordinating the presentation of patient profile data alongside reference ranges.

\begin{figure}[!htb]
    \centering
    \includegraphics[width=0.8\linewidth]{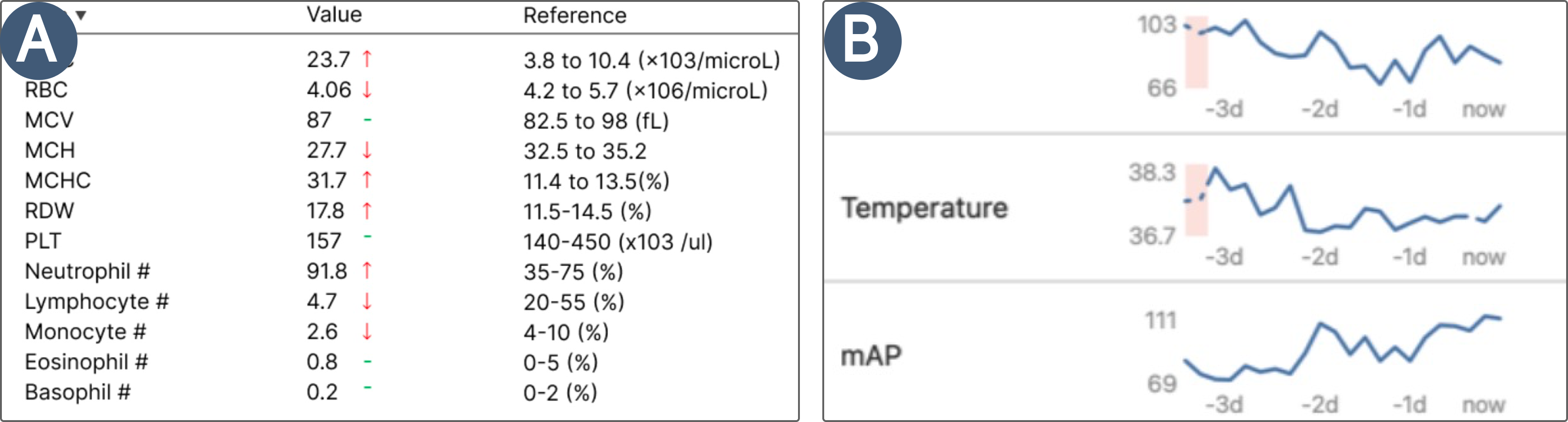}
    \caption{(A) Zhang et al.~\citep{Zhang2024Rethinking} display patient's profiles alongside reference ranges in a view for comparison. (B) Sivaraman et al.~\citep{Sivaraman2023Ignore} highlight abnormal areas directly on the patient trajectory.}
    \label{Pattern6}
\end{figure}

\noindent
$\blacklozenge$ \textit{The solution:} To address this problem, designers can display an individual patient's data alongside the corresponding established reference ranges, either side-by-side or overlaid.
\autoref{Pattern6}-A illustrates this pattern by juxtaposing them, while \autoref{Pattern6}-B highlights abnormal areas directly on the patient trajectory. 

\noindent
$\blacklozenge$ \textit{Justifications:} Similar to the former pattern, this presentation approach can enhance user performance in deriving insights across various dimensions of a feature without increasing their memory load. Through our interview, we also found that this design pattern aligns with the typical practices of health professionals.

\section{Information Presentation Interactivity}
\label{sec:info-org}
Interactions among different information entities are a crucial aspect of system design, including their order of presentation over time and the interactive coordination approaches between them. We identified six design patterns for interactivity.

\subsection{Target Patient Analysis}
\pattern{icon/o1}{Delayed Prediction}

\noindent
$\blacklozenge$ \textit{The problem:} 
If AI predictions are presented to health professionals before they formulate their own, they may be influenced by these predictions, leading to what is known as \textit{``confirmation bias''}~\citep{Bach2023If}.
Users need to overcome confirmation bias in high-stakes scenarios, as it may affect the accuracy of final outcomes if the AI predictions are incorrect.
The design problem is presenting AI predictions to mitigate potential bias.

\noindent
$\blacklozenge$ \textit{The solution:} To address this problem, designers can ensure that AI predictions are not displayed until users have made their preliminary choices, such as requiring users to input their decisions into the interface before displaying the AI suggestions. For example, To mitigate ophthalmologists' susceptibility to potential bias in detecting lesions in patients' eyes, AI results will only be displayed for evaluation after ophthalmologists complete their own assessments and submit them to the system~\citep{Bach2023If}. 

\noindent
$\blacklozenge$ \textit{Justifications:} 
Presenting AI output after users make initial decisions encourages reliance on their expertise before considering AI recommendations. In healthcare, where professionals possess significant knowledge, this method allows for independent evaluations, leading to a more objective assessment of AI outputs.
Through our interviews, nine participants (E3, E4, E5, E6, E7, E9, E10, E11, and E12) highlighted it may be more beneficial for novices. 
For instance, E3 mentioned, \textit{``As a clinician with only one year of experience, if I am told that this model has high accuracy, seeing the AI decisions might cause me to make premature assumptions.''}
Furthermore, we need to consider the urgency of time. 
For example, three participants (E2, E10, and E11) noted that in urgent situations, such as predicting sepsis, if the AI indicates a high-risk prediction, it should be presented to clinicians immediately for evaluation, ensuring a more effective response to the situation.
Therefore, designers should consider two key points: 1) the potential for confirmation bias, especially among less experienced professionals, and 2) the time factor, as some users may feel this approach could prolong decision-making.

\pattern{icon/o2}{Hiding Explanations by Default} 

\noindent
$\blacklozenge$ \textit{The problem:} AI explanations can sometimes be complex and may be unfamiliar to health professionals, potentially leading to information overload.
Health professionals need to manage their time effectively and may not be able to review AI explanations for every case.
The design problem is to prioritize the presentation of critical information, ensuring that complex AI explanations do not distract users.

\noindent
$\blacklozenge$ \textit{The solution:} To address this problem, designers can hide AI explanation information by default on the human-AI interface. Instead, users can access these explanations on demand, typically when their decisions differ from those of the AI. 
For instance, Gu et al.~\citep{Gu2023Augmenting} designed an explanation card that includes feature attributions of a tumor image, allowing pathologists to examine it as needed.

\noindent
$\blacklozenge$ \textit{Justifications:} 
Currently, multiple systems allow users to only check the explanations on demand---often when their own results differ from the AI's decisions---to ensure efficiency~\citep{Krause2016Interacting, Xie2020CheXplain, Francisco2021Breast, Gu2023Augmenting}.
Considering that information entities related to explanations might not be familiar to users, which can increase cognitive load, it is important to present explanations at appropriate times. 
When users and AI are in agreement, the AI outcomes are often sufficient to serve as confirmation of human decisions, eliminating the need for additional explanations.
However, when users and AI disagree, explanations can be used \textit{``as confirmatory''}~\citep{Gu2023Augmenting} to help users assess their judgment. 
This pattern enhances user efficiency and reduces their information burden.
However, when using this pattern, designers are advised to consider the format of the explanations. For example, two participants (E4 and E8) stated that saliency maps are not difficult for them to understand, and displaying them alongside AI predictions may not result in significant information overload.

\pattern{icon/o3}{What-if Analysis} 

\noindent
$\blacklozenge$ \textit{The problem:} 
Health professionals need to evaluate how changes in treatment plans or patient profiles could impact patient health outcomes or recovery. The design problem is how to enable users to simulate different input features and view the corresponding prediction results.

\noindent
$\blacklozenge$ \textit{The solution:} To address this problem, designers can create dynamic connections between AI-generated predictions and specific patient data.  
For example, Zhang et al.~\citep{Zhang2024Rethinking} allows users to select different lab test types and set their values to observe changes in predicted risk levels with confidence.


\noindent
$\blacklozenge$ \textit{Justifications:} 
This pattern has been leveraged in multiple systems, such as systems for risk screening~\citep{Kwon2019RetainVis}, clinical diagnosis~\citep{Cheng2022Vbridge, Ayobi2023Computational, Yin2024SepsisLab, Zhang2024Rethinking, Krause2016Interacting}, and pathological diagnosis~\citep{Cai2019Human-Centered}. Linking AI output interactively with patient profiles allows stakeholders to simulate various scenarios and assess the potential impact of different treatment options or changes in patient conditions, enhancing personalized medicine (E4, E6, E10, and E11).
For instance, E11 remarked, \textit{``When considering treatment for a patient with hypertension, I can adjust the types of medications, dosages, and combinations to see how these changes affect the patient’s blood pressure.''} 
However, designers must consider feature correlations in what-if analyses. For instance, when ``blood pressure'' increases, ``heart rate'' often rises as well. Adjusting only ``blood pressure'' while keeping ``heart rate'' constant may lead to inaccurate results. 

\pattern{icon/o4}{Input-Explanation Visual Linking}

\noindent
$\blacklozenge$ \textit{The problem:}
Feature attributions can help healthcare professionals quickly identify abnormal features detected by AI. They can also serve as a confirmation tool when users identify features they find notable.
Therefore, users need to conduct a \textit{``back-and-forth''}~\citep{Cheng2022Vbridge} analysis to identify and interpret noteworthy features, which involves switching between feature attributions and patient trajectories.
The design problem is integrating feature attributions and patient trajectories for comparative analysis.

\begin{figure}[!htb]
    \centering
    \includegraphics[width=0.6\linewidth]{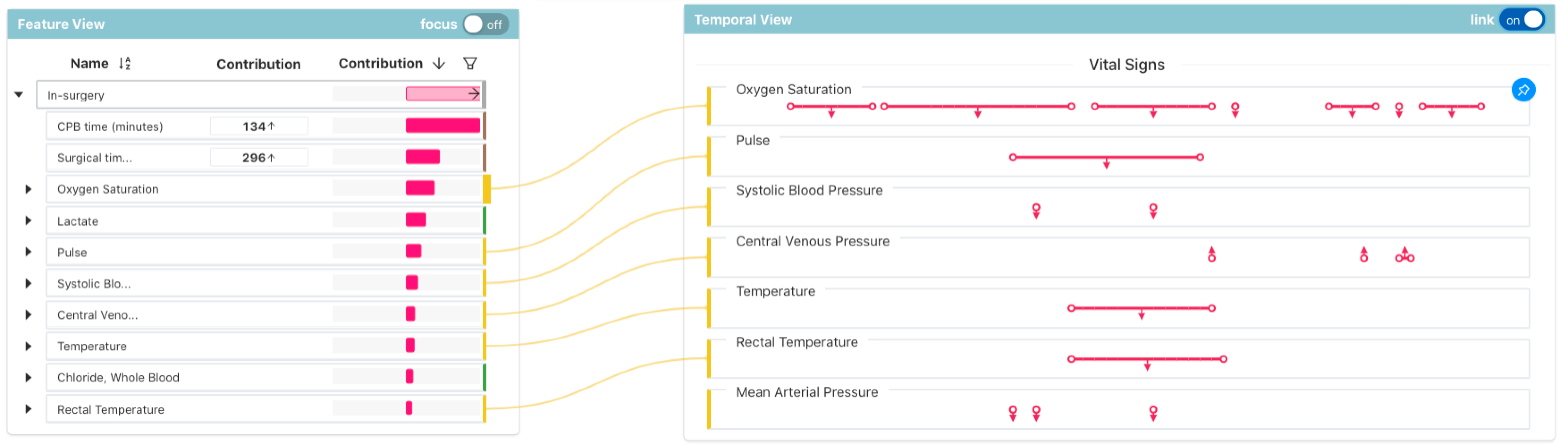}
    \caption{VBridge~\citep{Cheng2022Vbridge} associates feature attributions with corresponding patient trajectory by visual linking.}
    \label{Pattern10} 
\end{figure}

\noindent
$\blacklozenge$ \textit{The solution:}
To address this problem, designers can visually connect a feature attribution to its corresponding patient’s trajectory such as lab tests or vital signs.
\autoref{Pattern10} illustrates this pattern by enabling clinicians to trace the corresponding original data (\ie, patient trajectory) based on visual linking when they observe a high attribution value for a feature.

\noindent
$\blacklozenge$ \textit{Justifications:} 
Visually linking the two information entities can make users \textit{``easily get connections''}~\citep{Cheng2022Vbridge} between them by enhancing spatial consistency.
Furthermore, this pattern also guarantees simplicity by presenting different information in separate views, ensuring clarity (E6, E7, and E10).
However, during use, it is important to consider users' visualization literacy. They may not fully understand the meaning of visual linking, which necessitates prior user onboarding (E5 and E11).

\subsection{Relevant Cohort Exploration}
\pattern{icon/o5}{Interactive Cohort Selection} 

\noindent
$\blacklozenge$ \textit{The problem:} Users need to select appropriate patient cohorts to calculate feature or outcome statistics as references. 
Besides, they sometimes need to narrow their focus when selecting specific patients for inspection.
The design problem is how to present patient cohorts for effective exploration.

\begin{figure}[!htb]
    \centering
    \includegraphics[width=0.3\linewidth]{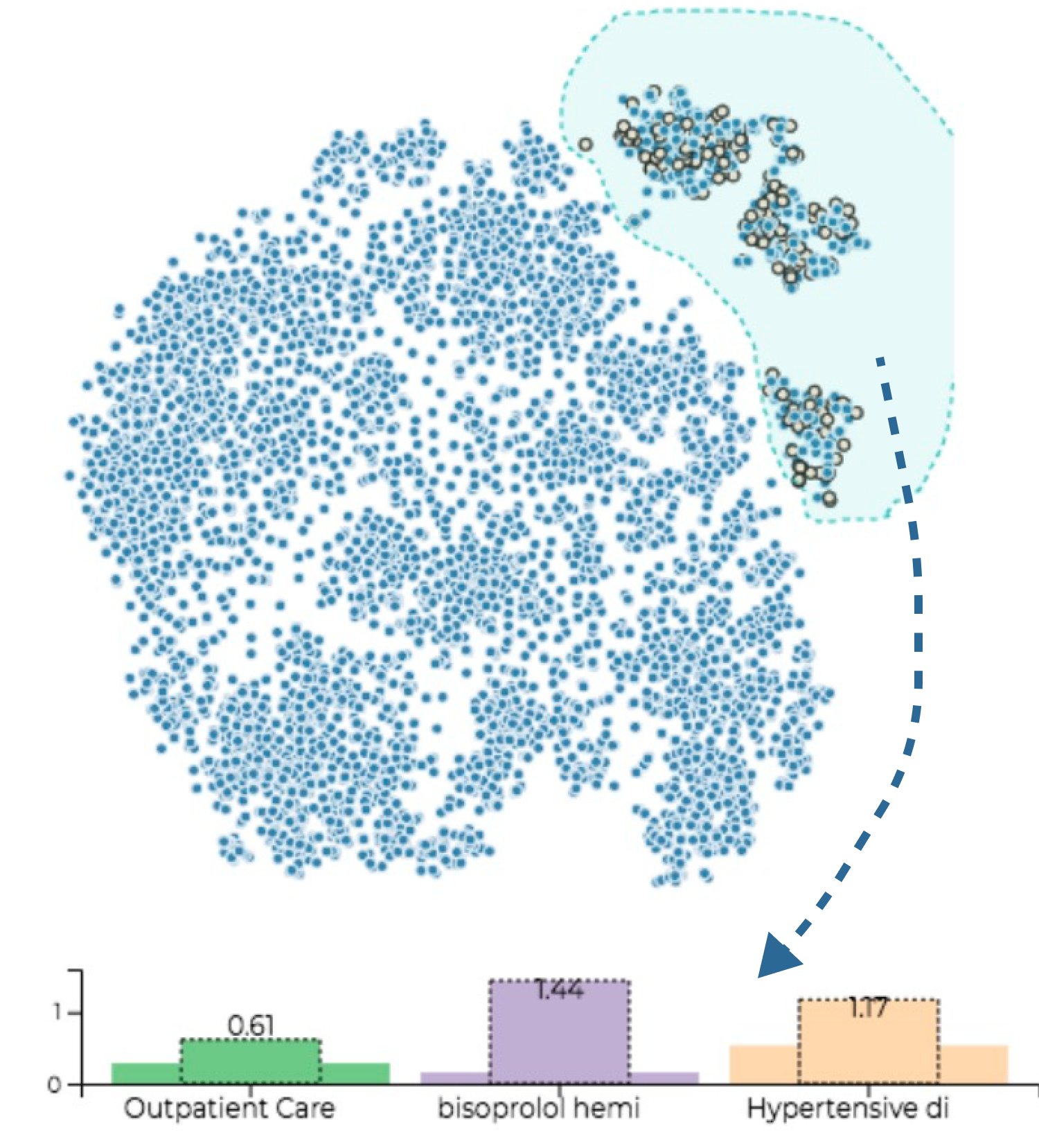}
    \caption{RetainVis~\citep{Kwon2019RetainVis} enables users to select patients through an overview and examine their corresponding cohort statistics.}
    \label{Pattern11} 
\end{figure}

\noindent
$\blacklozenge$ \textit{The solution:}
To address this problem, designers can enable users to explore cohort statistics based on an overview in an interactive manner.
\autoref{Pattern11} demonstrates this pattern using a scatter plot to display an overview of patients, enabling users to lasso a specific region and dynamically update the cohort statistics.

\noindent
$\blacklozenge$ \textit{Justifications:} 
Firstly, allowing users to select relevant patients from the cohort overview can integrate users' domain knowledge, which leads to more personalized care (E4, E5, and E12).
Furthermore, this interactive approach is well-suited for iterative exploration, enabling users to refine their analyses as they uncover new patterns and insights (E1, E3, E5, E6, E9, and E10). 
However, ensuring that users can accurately select an appropriate patient group for analysis is a key consideration for designers (E4 and E10).
For example, designers can use color, size, or other visual marks to encode different patients in the overview, providing users with more information to help them make accurate selections.

\subsection{Incorporation of External Knowledge or Metadata}
\pattern{icon/o6}{Metadata on Demand} 

\noindent
$\blacklozenge$ \textit{The problem:}
A lot of metadata, such as the sample size of the dataset and the specific models used, may not be critical for health professionals when making decisions. Additionally, this information might be quite complex and overwhelm users.
The design problem is to enable health professionals to focus on the most important information (such as input features or AI predictions) and avoid becoming distracted by excessive metadata.

\noindent
$\blacklozenge$ \textit{The solution:}
To address this problem, additional metadata information can be provided only when users actively engage by clicking on a trigger. For instance, xOpat~\citep{Horák2023xOpat} hides the implementation details of the system initially and provides a button that allows users to view this information when needed.

\noindent
$\blacklozenge$ \textit{Justifications:} 
This pattern has been implemented in multiple existing systems~\citep{Francisco2021Breast, Cheng2022Vbridge, gomez2023designing, Horák2023xOpat, Wentzel2025DITTO}. Presenting metadata in a popover component after clicking not only provides easy access to important contextual information---such as model validation steps and results~\citep{Jacobs2021Design}---but also respects the limited time that stakeholders have with patients.
For instance, E10 mentioned that \textit{``Reviewing the details of the dataset only makes me lose track in a sea of unfamiliar information. This pattern can help to mitigate this problem.''}
When designers use this pattern, it is advisable to differentiate between types of metadata.
For instance, critical metadata, such as the source of the training data (whether it comes from their own hospital or other renowned hospitals), might be displayed directly (E4, E6, and E12). In contrast, other metadata, like model details, can be presented upon clicking.
\section{Workshop}
\begin{table}[ht]
\sidecomment{R4C7}
\centering
\caption{\rui{Demographics of the 14 UI designers in our workshop.}}
\label{table2}
\begin{tabular}{l|l|l|l|l|l}
\hline
\textbf{ID} & \textbf{Gender} & \textbf{Experience}   & \textbf{Specialization} & \textbf{K(AI)} & \textbf{K(Med)} \\ \hline
P1          & Male            & 3-5 years             & Research                & 3                      & 3                          \\ \hline
P2          & Male            & 1-3 years             & Research                & 2                      & 2                          \\ \hline
P3          & Female          & \textless{}1 year     & Industry                & 3                      & 1                          \\ \hline
P4          & Female          & 3-5 years             & Industry \& Research    & 4                      & 2                          \\ \hline
P5          & Male            & 1-3 years             & Industry \& Research    & 3                      & 2                          \\ \hline
P6          & Male            & \textless{}1 year     & Research                & 2                      & 1                          \\ \hline
P7          & Male            & \textgreater{}5 years & Research                & 5                      & 2                          \\ \hline
P8          & Male            & 1-3 years             & Research                & 2                      & 2                          \\ \hline
P9          & Male            & 3-5 years             & Research                & 2                      & 2                          \\ \hline
P10         & Female          & 3-5 years             & Research                & 1                      & 3                          \\ \hline
P11         & Female          & 1-3 years             & Research                & 5                      & 1                           \\ \hline
P12         & Male            & \textgreater 5 years  & Industry                & 2                      & 2                          \\ \hline
P13         & Male            & \textless{}1 year     & Research                & 1                      & 1                          \\ \hline
P14         & Male            & 3-5 years             & Research                & 3                      & 3                          \\ \hline
\end{tabular}
\end{table}
The main goal of the workshops is to gather participants' usage strategies and reflections on our design patterns. 
\sidecomment{R3C9}
\rui{We recruited participants through online networks, including Xiaohongshu, Weibo, and X.}
The primary recruitment criterion was whether they had previously designed human-AI systems for health professionals. 
Additionally, they needed to be willing to share their own design contexts during the workshop.
Ultimately, we recruited 14 participants (P1-P14) with experience in designing healthcare human-AI interfaces across different scenarios (\autoref{table2}).
Each workshop session included 2 to 4 participants. We began with a 30-minute introduction to our design patterns, followed by 20 minutes for participants to familiarize themselves with them. Next, participants presented their own design contexts related to our scope for 30 minutes.
They then created an interface sketch of a healthcare human-AI system in Miro using our design patterns for 40 minutes, followed by 10 minutes of peer discussion. Finally, they presented their sketches and reflected on their design choices and challenges for 20 minutes. This process aimed to gather usage strategies and identify future opportunities.
Each workshop session lasted for around 2.5 hours. The output of the workshop includes 14 interface sketches. 
The workshop was approved by the IRB, and we obtained consent to share their design sketches and intermediate outputs, ensuring no materials were included in our paper without permission.



\subsection{Workshop Findings} \label{findings}

\sidecomment{R4C7}
\rui{To analyze the workshop data, we applied thematic analysis to the transcripts, notes, and sketching artifacts. 
Two authors independently coded participants’ comments and observations to identify recurring insights.
Coding focused on usage strategies of the design patterns and their reflections on the advantages and limitations of the patterns. Discrepancies in coding were resolved through discussion until consensus was reached.}

\subsubsection{Design Pattern Usage Strategies} 
\sidecomment{R4C8}
\rui{We identified four scenarios in which users would utilize our design patterns: understanding user needs, exploring more design options, simplifying the interface, and inspecting unarticulated needs.}

\rui{\textit{\textbf{Grounding user needs using information entities.}} 
Firstly, we observed that all participants would encounter abstract user needs and tried to translate these user needs into concrete information entities or design patterns to better interpret and fulfill user needs.
Some participants would decompose those user needs (e.g., ``help me understand why the AI disagrees with me'') into concrete functional needs, often represented as relationships among different types of information (e.g., comparison between human and AI reasoning).
The information they identified did not always align perfectly with our predefined information entities, so they subsequently created mappings—for example, linking AI reasoning to entities such as ``case-based explanations'' and ``literature evidence''.
P13 mentioned, \textit{``At this point, user needs become very concrete and are easily understood and fulfilled.''}
In addition, the documentation we provided also helps them understand medical terms in user needs. For example, P6 identified the information entity ``Feature Attributions'' as suitable for addressing the clinician’s need about ``I would like to know which symptoms are most critical for making this decision''.}

\rui{\textbf{\textit{Exploring more design options.}} Next, we found seven (of 14) participants, when addressing a specific user need, aimed for a comprehensive view of design options. They mentioned that this not only enhanced the optimization of their designs but also improved their confidence.
Specifically, they would identify all potentially relevant information entities and design patterns, and then comprehensively compare them for more informed design choice.
When participants were familiar with the medical scenario, they could make decisions directly by consulting our documentation, which summarized the advantages and considerations of each design pattern. 
Otherwise, they considered these options as candidates to discuss with domain experts.
Moreover, we observed that our participants did not simply select among alternatives—they also considered whether different options could be integrated.
For example, when trying to support clinicians in understanding which actions should be taken for a particular symptom, P11 considered several alternatives: showing similar past cases, illustrating the effects of different actions on outcomes using partial dependence plots, and enabling interactive scenario exploration through what-if analysis. 
She evaluated these options in terms of transparency, informativeness, and how easily a clinician could interpret the results. Ultimately, she chose to combine the patterns of ``Composed PDPs'' with ``What-if Analysis'': the PDPs provided an overview of how varying actions would change outcomes, while the what-if analysis allowed clinicians to interactively explore alternative decisions and reason about specific scenarios. 
As P11 reflected, \textit{``What mainly helped me consider combining these two patterns was that they involved overlapping information entities, which inspired me to integrate them.''}
}

\rui{\textbf{\textit{Simplifying the interface.}} Then, we observed that seven (of 14) participants simplified the interface using our design patterns to reduce users' cognitive load after completing their designs based on prior knowledge.
These participants faced challenges related to information completeness and cognitive load: presenting all relevant data at once ensured nothing was missing but created clutter and increased interpretive effort for clinicians.
For instance, P7 initially displayed all patient data, AI predictions, and supporting evidence simultaneously. While comprehensive, this approach made it difficult for users to focus on the most relevant insights.
Then he started to consult relevant patterns such as ``Hiding Explanations by Default'' and ``Metadata on Demand'', he was able to simplify the presentation while preserving essential content.
This process illustrates a deliberate design strategy: participants used their prior knowledge to generate an initial, holistic sketch, then applied design patterns to optimize the balance between completeness and cognitive manageability.
}

\rui{\textbf{\textit{Identifying unarticulated needs.}} Finally, we found that five (of 14) participants reviewed relevant design patterns based on the information entities they had already included, using these patterns to inspect overlooked aspects for refinement.
They were concerned that some user needs might not have been explicitly articulated by experts but could still affect the system's effectiveness.
We observed that these unarticulated user needs could stem from designers’ limited understanding of medical conventions—such as presenting a patient’s original data alongside references—or from overlooking human cognitive risks, for example, potential biases introduced by AI predictions.
P14 noted, \textit{``These design patterns serve well as cues, indicating what previous systems have already considered.''}
While one approach was to systematically check all design patterns, we also observed another method: participants centered their review on the specific information entities in their sketches and considered relevant deign patterns.
For example, P3, after presenting the patient’s original data, noticed the \textit{``Patient with References''} pattern. Although clinicians had not explicitly mentioned this aspect—perhaps because they assumed it was obvious—P3, with less domain-specific knowledge, recognized it as a valuable addition that enhanced both the completeness and clarity of the interface.
}

\subsubsection{User Perceptions of Our Design Patterns} 
Firstly, five (of 14) participants mentioned that those design patterns could assist them in \textit{\textbf{initial discussions with stakeholders for design requirement analysis}}.
Specifically, prior to creating design sketches in the workshop, many participants noted that they faced significant challenges in efficiently gathering user requirements.
For example, P3 mentioned, \textit{``Often, when we conduct interviews to derive user requirements, we hope stakeholders will directly tell us what information they want to examine. The problem is that clinicians may struggle to articulate these specifics because they themselves may not fully understand what they want.''} 
Additionally, several participants highlighted that due to differences in domain knowledge, their interpretations of clinicians’ comments may not accurately reflect their true needs.
P5 shared, \textit{``I previously worked on helping clinicians better assess the accuracy of AI decisions. When clinicians expressed a desire to know what evidence supports or contradicts AI decisions, I showcased feature attributions. However, they indicated that this was not in line with their traditional processes and preferred literature evidence instead.”}
Equipped with our design patterns, designers could present these options directly to clinicians, enabling them to make choices rather than providing open-ended responses, which can enhance communication efficiency.

Secondly, 
\sidecomment{R3C3}
\rui{all the participants appreciated that our design patterns have \textit{\textbf{provided concrete design solutions}}}.
P13 mentioned, \textit{``Users often express very abstract needs, such as viewing the importance of information. Then we will brainstorm design ideas based on these abstract requirements. However, many times, our thoughts do not align with what technical personnel find practical.''}
Our design patterns encourage starting with information entities, offering clear guidelines for initiating the human-AI interface design process. 
\sidecomment{R3C3}
\rui{Compared with previous general design principles, this information-driven approach provides designers with actionable and concrete methods to leverage available data and better address user needs.}
Additionally, seven participants noted that our design patterns helped them better organize user needs from two aspects (\ie, presentation coordination and interaction).
For example, P14 highlighted, \textit{``Some clinicians tend to jump from one topic to another when elaborating on their requirements, sometimes discussing what information they require and other times focusing on how they want to view it.''} 
Our design patterns provide a systematic approach to classify and map these different user needs to specific designs.

Thirdly, four (of 14) participants praised our \textit{\textbf{block-building approach to summarize design patterns}}, which involves first constructing information entities and then using these entities to develop various design patterns. This method is not only easy to understand (P2 and P6) but also encourages designers to consider the potential for new solutions. 
Specifically, P9 mentioned that our pattern of ``Patient vs Cohort'' sparked his thoughts on AI explanations. He observed that while the examples we provided only included input features or outcomes for comparison with the cohort, ``Feature Attributions'' and ``Global Feature Importance'' essentially represent the relationship between the target and the cohort. Comparing these two types of explanations could help clinicians understand patient specificity. P9 stated, \textit{``For example, while global importance may highlight that oxygen saturation is crucial, feature attributions might reveal that blood glucose levels have the largest impact. This could assist clinicians in better understanding both the general characteristics of the disease and the individual’s uniqueness.''}

Fourthly, three (of 14) participants mentioned that our design patterns could help them \textit{\textbf{increase awareness of cognitive aspects of stakeholders}}.
P2 stated that when designing interfaces, they often overlooked users' cognitive load, leading them to focus more on whether information is presented.
\textit{``Given the vast amount of information involved in AI systems, organizing and presenting this information effectively is crucial. The information coordination patterns greatly help to mitigate visual clutter and reduce cognitive load.''}
Furthermore, P4 and P5 noted that our patterns could enable them to pay attention to biases introduced by the information.
For instance, P4 stated, \textit{``I had always focused too much on visualizing all information in an interface. However, the design patterns related to presentation order reminded me to consider human bias during the high-stakes decision-making process.''}
P4 also thought that these patterns could be effective not only in designing health-related interfaces but also in inspiring the design of interfaces in other contexts like AI-assisted resume screening.

\subsection{Case Study}

\begin{figure}[t]
    \centering
    \includegraphics[width=1\linewidth]{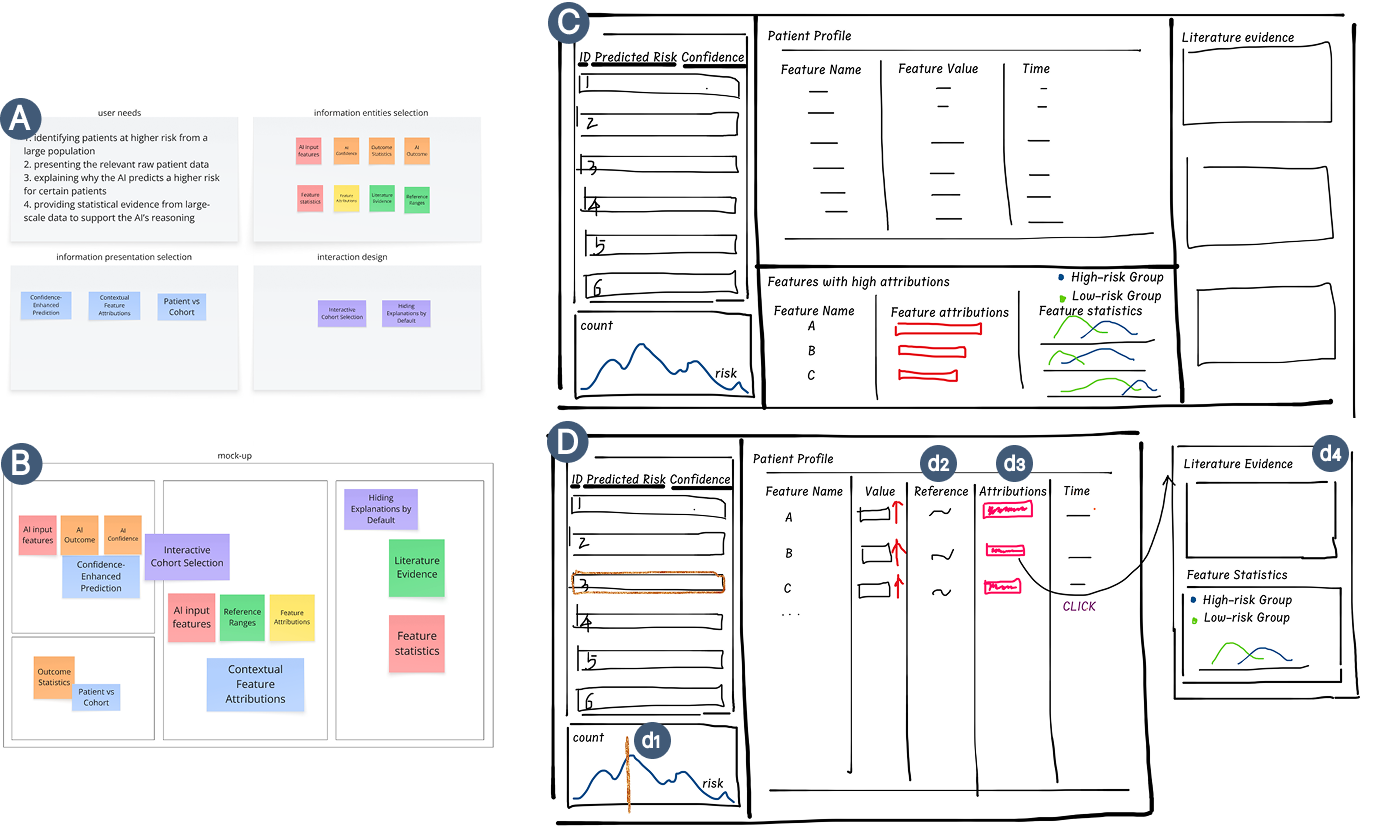}
    \caption{(A) The stickers that illustrate the selected information entities and design patterns based on the user needs. (B) The sketch that utilizes the stickers for presentation. (C) The interface sketch of the initial design. (D) The interface sketch of the final design that reflects refinements made using the 12 design patterns.}
    \label{sketch} 
\end{figure}

\sidecomment{R3C12, R1C5\_1}
\rui{To illustrate the application of our design patterns, we selected an example from our workshop with participant consent. Specifically, the designer Jack (a pseudonym) was tasked with creating a system to monitor potential postoperative complication risks, guided by four user needs: \textbf{(T1)} identifying patients at higher risk, \textbf{(T2)} presenting the relevant raw patient data, \textbf{(T3)} explaining why the AI predicts a higher risk, and \textbf{(T4)} providing statistical evidence to support the AI’s reasoning.
}

\rui{Jack, with limited medical and AI expertise, mapped these needs to information entities (\autoref{sketch}-A).
To satisfy \textbf{T1}, Jack relied on the \textit{AI outcome} and \textit{AI confidence} as primary screening entities. Then he noticed the \textit{outcome statistics} and thought, \textit{``This might actually give clinicians the population context—help them understand what the average risk looks like.''}
To address \textbf{T2}, he decided to present the patient attributes and medical history for each flagged patient. Therefore, he selected the \textit{AI input features} for this need.
For \textbf{T3}, he noticed there were many information entities available, such as \textit{feature attributions} and \textit{case-based explanations}. He stated, \textit{``Clinicians often make judgments based on abnormal symptoms and their severity...maybe feature attributions would be more intuitive for them.''}
Finally, for \textbf{T4}, Jack realized this goal was less concrete. He recalled clinicians mentioning that if the AI flagged a patient due to certain symptoms, it would be helpful to know whether large-scale past cases supported this reasoning. He thought, \textit{If the AI and the clinician disagree, having supporting evidence could really help guide the judgment.''} Carefully reviewing the available information entities, he identified \textit{feature statistics} as potentially useful. Interestingly, he also discovered \textit{literature evidence}, thinking, \textit{``If we can find papers that support or refute the AI’s reasoning, it would be extremely helpful.''}
}

\rui{Based on his design experience, Jack sketched an initial layout using the stickers. 
He then created a draft sketch (\autoref{sketch}-C).
The interface displays each patient’s ID, AI-predicted risk score, and AI confidence in a list format, with an area chart showing overall prediction statistics. This layout helps clinicians prioritize patients for further inspection. Selecting a patient takes users to a middle view with patient profiles and a lower view highlighting high-attribution features and their statistics. In the final view, relevant literature is retrieved to provide additional support or counter-evidence for the AI’s reasoning.}

\rui{Next, he began optimizing the sketch.
The design outcome using the stickers can be seen in \autoref{sketch}-B, featuring a more detailed version (\autoref{sketch}-D).
He first focused on those design patterns where all the involved information entities were present in the interface.
He noticed the ``Contextual Feature Attributions'' pattern. 
Initially, he displayed high-attribution features separately, which led to redundancy. To improve clarity, he placed feature attributions alongside their values (\autoref{sketch}-d3).
He also considered the ``Patient vs. Cohort'' pattern (\autoref{sketch}-d1). 
For the feature statistics and literature evidence, he drew inspiration from the ``Hiding Explanations by Default'' pattern, adding interactive elements that allow clinicians to access detailed statistics only when needed (\autoref{sketch}-d4). This round of optimization made the system more streamlined and visually concise.}

\rui{Finally, Jack used the design patterns as a checklist to ensure that no important design aspects were overlooked. 
Specifically, he examined those design patterns that involved any of the information entities already present in his interface.
He identified the ``Patient with References'' pattern as particularly important and added an additional column to display the corresponding reference ranges (\autoref{sketch}-d2).
After ensuring that no further design considerations needed attention, he finalized the design sketch.}
\section{Discussion}

In this section, we discuss the implications of this study on human-AI interface designs, the generalizability of the proposed design patterns for other domains, and the limitations of this study.

\subsection{Design Implications}

\sidecomment{R1C4\_1, R1C5\_3}
\rui{\textbf{Reusing the Clinical Human-AI Interface Designs.}
Designing human-AI interfaces in healthcare is particularly challenging due to the reliance on deep, domain-specific knowledge, much of which is implicit and inaccessible to designers. We address this challenge by distilling recurring solutions from historical designs into common design patterns.
Specifically, we extracted information entities as the foundation of our design patterns.
This approach not only aligns with the characteristics of AI but also provides designers with a step-by-step method for utilizing our patterns—beginning with user needs to determine which information entities to display. Additionally, this method offers flexibility for future expansions, allowing new information entities and design patterns to be seamlessly integrated into the existing framework.
In the future, as additional clinical conventions are identified, our framework can be extended to encode a wider range of medical reasoning processes into reusable interface patterns.
}

\textbf{Towards a Design Language for Human-AI Interfaces.}
A design language system, like Apple Human Interface Guidelines (HIG)~\footnote{https://developer.apple.com/design/} and IBM Design Language~\footnote{https://www.ibm.com/design/language/}, usually includes a collection of design elements, patterns, and guidelines that inform the design process.
In this paper, we make an initial approach toward a design language system for human-AI interfaces by categorizing the information entities and summarizing the design patterns in information presentation coordination and interactivity. 
There are more open questions regarding \textit{human actions and inputs to the model}, \textit{user onboarding}, and \textit{accessibility supports} in human-AI interfaces.
For example, when summarizing design patterns considering accessibility support, we can place greater emphasis on the use of color in information presentation and consider layout adjustments to enhance visibility and usability.


\textbf{Providing Design Guidance for Data-Driven Interfaces.}
Unlike conventional UI design patterns that describe general UI component designs, our paper focuses on the information (or data) entities in healthcare human-AI interfaces and patterns in presenting and arranging these information entities.
This focus is important because human-AI interfaces are typically driven by data.
In this situation, designers will face new challenges compared with conventional UIs.
For instance, designers need to consider how to select data entities and create proper visual representations.
To address those challenges, we developed design patterns specifically for health-related human-AI interfaces.
When summarizing the design patterns, we borrow the taxonomy of elementary visualization and composite visualization types~\citep{Javed2012Composition} from the visualization community. 
The design patterns from our paper can complement conventional UI patterns. 
The two design pattern systems could be used together in designing human-AI interfaces in healthcare.

\textbf{Implications for Summarizing Design Patterns.}
We emphasize the importance of incorporating two user groups when analyzing and compiling design patterns: one being UI designers and the other being end users (\ie, health professionals in our context). Compared to previous works that focused on only one group---either UI designers or end users~\citep{Schoonderwoerd2021Human-centered, Bach2023Dashboard, Bach2018Comics}---engaging both groups allows for more comprehensive actionable insights.
Firstly, since the designed systems are ultimately used by end users such as clinicians and pathologists, their perspectives on design patterns are crucial. Additionally, as our design patterns are intended for direct use by UI designers, understanding how they apply these patterns and how the patterns support their work is essential. This insight can help designers effectively implement these patterns in real-world applications.

\sidecomment{R4C9, R3C5}
\rui{\textbf{Design Patterns as Communication Artifacts in Co-Design.}
During the workshop, participants indicated that our design patterns could facilitate co-design between designers and medical professionals. 
They help clinicians articulate needs by grounding discussions in concrete information entities relevant to the medical context. At the same time, these design patterns assist designers in translating clinicians' abstract needs into specific design solutions.
This mutual enhancement fosters a more effective co-design process.
Future work should explore how these patterns can guide co-design discussions, potentially embedding them into conversation flows for autonomous recommendations that help medical professionals articulate their needs. Additionally, these patterns could drive prototype generation rather than serving solely as references, enabling clinicians to select patterns and automatically create corresponding interactive modules. 
}

\subsection{Generalizability of Our Work.}
Although our design patterns are targeted at the healthcare domain, they may also be applicable in other fields by (1) revising the information entities to be more tailored to the specific domain, and (2) abstracting the data types of the information entities in our design patterns to align with the relevant data in the specific domain.

We use the design of AI-assisted interfaces for climate risk prediction as one example.
First, numerous information entities we have identified can still be applied, such as different types of explanations. 
However, end users in the environment domain might require other external knowledge to help them judge AI decisions, such as simulation results of traditional numerical models or news data~\citep{li2024Save}.
Furthermore, design patterns related to information presentation coordination require transforming healthcare information into environmental data. A simple approach is to use data types as a mapping abstraction.
For example, the information in the pattern of juxtaposing patient trajectories with forecasting results can be generalized to two types of time series data (\ie, historical and predicted time series).
This can align with the variation and forecasting of air quality over time in the environmental domain.
Therefore, designers can further consider this pattern in such cases.
Finally, it is crucial to revise and summarize the design justifications for each pattern based on the environment domain requirements.

\sidecomment{R1C4\_4, R1C5\_4}
\rui{Similarly, our design patterns can be adapted for stock market prediction, where human-AI systems assist users in forecasting stock trends or recommending investment strategies. Several design patterns. For instance, ``Confidence-Enhanced Prediction'' is critical for assessing risk and making informed investment decisions.
Similarly to the previous example, we can abstract data types to suit the stock market domain (e.g., historical stock data, market trends). For instance, we can adapt ``Trajectory-Forecast Juxtaposition'' to compare stock performance over time and project future price trends. 
However, we should also note that several design patterns may not be as applicable. For instance, ``Delayed Prediction'', may be less relevant in the fast-paced stock market, where timely decision-making is critical.}

\subsection{Limitations and Future Works.}
\textbf{Reliance on Literature for Pattern Identification.}
One limitation is that we focused solely on extracting design patterns from existing literature rather than from currently deployed human-AI systems in the real world. The primary reason for this choice is that these systems are not always accessible, unlike published research papers. In addition, we wanted to understand the rationales behind designers' decisions, which are well-documented in the literature.
\sidecomment{R3C5, R1C5\_5}
\rui{
Future work could explore the applicability of the design patterns proposed in our paper by comparing them with deployed systems and interviewing stakeholders who have experience using such systems.}

\textbf{Applicability in Various Healthcare Scenarios.}
The papers we collected are primarily divided into four scenarios, including risk screening, clinical diagnosis, treatment planning, and pathological diagnosis. We acknowledge that these scenarios cannot fully capture the diverse applications of AI in healthcare like remote patient monitoring.
\sidecomment{R3C5, R1C4\_2, R1C5\_5}
\rui{Moreover, our design patterns have not yet been categorized according to critical factors like time urgency or shared decision-making with patients. For instance, high-urgency contexts, such as trauma care, require rapid, clear recommendations with minimal interaction, while non-urgent settings, like preventive care, allow for more in-depth clinician engagement. In shared decision-making scenarios, common in chronic disease management or mental health care, AI tools must facilitate collaboration between clinicians and patients. These systems should be transparent, easy for patients to understand, and promote dialogue. Although our current design patterns provide justifications and considerations to guide designers on when to apply them, future work should address these factors to create more clearly categorized and context-sensitive solutions that better meet the diverse needs of healthcare users.}

\textbf{Leveraging LLMs for Adaptive Use of Design Patterns.}
\sidecomment{R3C5, R1C4\_3, R1C5\_5}
\rui{Our current design patterns are prescriptive and document-based, relying on designers to manually map user needs to information entities and interface solutions. With the emergence of Large Language Models (LLMs), the design process can become more dynamic and adaptive.
LLMs can act as collaborative design partners capable of interpreting design goals, contextualizing user needs, and recommending patterns most relevant to the situation at hand. Beyond simple retrieval, they can explain the rationale behind each pattern, compare alternative options, and even generate example interface variations, allowing designers to immediately evaluate and refine ideas in context.
Through such interaction, design patterns evolve from static references into generative instruments that actively facilitate exploration, reflection, and iteration. In this sense, LLMs have the potential to transform design patterns from fixed repositories of knowledge into adaptive tools that support continuous, human–AI co-design.}
\section{Conclusion}
Despite efforts in the HCI community to create design principles for human-AI interfaces, translating these guidelines into concrete designs is still challenging. This work presents a systematic guideline for designing healthcare-related human-AI interfaces, assisting users in three key steps: information entity selection, information entity presentation, and interaction design. Workshops with 14 participants evaluated our patterns, and feedback indicated our approach offers a more detailed, systematic, and data-driven perspective, marking a significant advancement in interface design.

\clearpage
\section*{Declaration of generative AI and AI-assisted technologies in the writing process.}

During the preparation of this work, the authors used ERNIE Bot in order to improve the readability and language of the manuscript. After using this tool, the authors reviewed and edited the content as needed and take full responsibility for the content of the published article.
\section*{Acknowledgement}

We would like to express our gratitude to all our collaborators for their invaluable assistance. We also acknowledge the support of the Hong Kong government. This work was partially supported by RGC GRF grant 16210722. Additionally, this research was also supported by the Hong Kong PhD Fellowship.

\clearpage
\bibliographystyle{elsarticle-num-names}
\bibliography{bib} 
\end{document}